\documentclass[12pt]{article}

\usepackage{fullpage}
\usepackage{amssymb}
\usepackage{bbm}
\usepackage{color}
\usepackage{ulem}
\usepackage[numbers,sort&compress]{natbib}

\usepackage{tocloft}

\newcommand{\Comment}[1]{{}}
\definecolor{darkblue}{rgb}{0.15,0.35,0.55}
\definecolor{reddish}{rgb}{0.65, 0.2, 0.2}
\definecolor{darkgreen}{RGB}{50,150,0}
\definecolor{greyish2}{rgb}{.96,.96,.96}
\usepackage[linktocpage=true]{hyperref}
\hypersetup{
colorlinks=true,
citecolor=darkblue,
linkcolor=reddish,
urlcolor=darkblue,
pdfauthor={},
pdftitle={},
pdfsubject={}
}

\DeclareFontFamily{OT1}{rsfs10}{}
\DeclareFontShape{OT1}{rsfs10}{m}{n}{ <-> rsfs10 }{}
\DeclareMathAlphabet{\mathscript}{OT1}{rsfs10}{m}{n}

\def\gsim{ \lower .75ex \hbox{$\sim$} \llap{\raise .27ex \hbox{$>$}} }
\def\lsim{ \lower .75ex \hbox{$\sim$} \llap{\raise .27ex \hbox{$<$}} }
\def\be{\begin{equation}}
\def\ee{\end{equation}}
\def\bea{\begin{eqnarray}}
\def\eea{\end{eqnarray}}

\newcommand{\ns}{\normalsize}

\newcommand{\rd}{{\rm d}}

\usepackage{latexsym,amsmath,amssymb,epsfig}

\usepackage{graphicx}
\usepackage{epstopdf}
\usepackage[body={17.5cm, 21cm},right=2cm]{geometry}
\usepackage{amssymb}
\usepackage{amsmath}
\usepackage{psfrag}
\usepackage{epsfig}
\usepackage{cancel}
 \allowdisplaybreaks[4]

\usepackage[all]{xy}

\begin{document}

\begin{titlepage}
\thispagestyle{empty}

\title{\vspace{-1.5cm}
  \hfill{\ns }  \\[1em]
   {\LARGE Non-linear Realizations of Conformal Symmetry and Effective Field Theory for the Pseudo-Conformal Universe}
\\[1em] }
\author{
   Kurt Hinterbichler, Austin Joyce and Justin Khoury
     \\[0.5em]
{\ns  Center for Particle Cosmology, Department of Physics and Astronomy,} \\[-0.4cm]
{\ns  University of Pennsylvania, Philadelphia, PA 19104}\\[0.3cm]
}
\date{}
\maketitle

\begin{abstract}
\noindent
The pseudo-conformal scenario is an alternative to inflation in which the early universe is described by an approximate conformal field theory on flat, Minkowski space.
Some fields acquire a time-dependent expectation value, which breaks the flat space $\mathfrak{so}(4,2)$ conformal algebra to its $\mathfrak{so}(4,1)$ de Sitter subalgebra. As a result, weight-0
fields acquire a scale invariant spectrum of perturbations. The scenario is very general, and its essential features are determined by the symmetry breaking pattern,
irrespective of the details of the underlying microphysics. In this paper, we apply the well-known coset technique to derive the most general effective lagrangian
describing the Goldstone field and matter fields, consistent with the assumed symmetries. The resulting action captures the low energy dynamics of
any pseudo-conformal realization, including the U(1)-invariant quartic model and the Galilean Genesis scenario.  We also derive this lagrangian using an alternative method
of curvature invariants, consisting of writing down geometric scalars in terms of the conformal mode. Using this general effective action, we compute
the two-point function for the Goldstone and a fiducial weight-0 field, as well as some sample three-point functions involving these fields.
\end{abstract}

\end{titlepage}

\setcounter{page}{2}
\tableofcontents

\newpage
\section{Introduction}
\numberwithin{equation}{section}
Cosmic microwave background and large-scale structure measurements provide strong observational evidence for a nearly scale invariant and gaussian spectrum of curvature perturbations in the very early universe. An important goal of early-universe cosmology is to understand the genesis of these fluctuations. The inflationary universe \cite{Starobinsky:1979ty, Guth:1980zm, Albrecht:1982wi, Linde:1981mu} addresses this question as well as the horizon and flatness problems, but it is not the unique mechanism by which to solve these problems.  This has led to proposed alternatives to the inflationary paradigm, for example, pre-big bang cosmology \cite{Gasperini:1992em, Gasperini:2002bn, Gasperini:2007vw}, string gas cosmology \cite{Brandenberger:1988aj, Nayeri:2005ck, Brandenberger:2006xi, Brandenberger:2006vv, Brandenberger:2006pr, Battefeld:2005av} and the ekpyrotic scenario~\cite{Khoury:2001wf, Donagi:2001fs, Khoury:2001bz, Khoury:2001zk, Lyth:2001pf, Brandenberger:2001bs, Steinhardt:2001st, Notari:2002yc, Finelli:2002we, Tsujikawa:2002qc, Gratton:2003pe, Tolley:2003nx, Craps:2003ai, Khoury:2003vb, Khoury:2003rt, Khoury:2004xi, Creminelli:2004jg, Lehners:2007ac, Buchbinder:2007ad, Buchbinder:2007tw, Buchbinder:2007at, Creminelli:2007aq, Koyama:2007mg, Koyama:2007ag, Lehners:2007wc, Lehners:2008my, Lehners:2009qu, Khoury:2009my, Khoury:2011ii, Joyce:2011ta}.

For a single scalar degree of freedom minimally coupled to Einstein gravity -- and with luminal or sub-luminal sound speed -- demanding that a solution both produces a scale-invariant spectrum of curvature perturbations on a dynamical attractor background and remains weakly-coupled over many decades of modes leads one uniquely to inflation \cite{Khoury:2010gw, Joyce:2011kh, Baumann:2011dt, Geshnizjani:2011dk}. Therefore, alternative mechanisms which generate perturbations while remaining weakly-coupled must either rely on an instability, as in the contracting matter-dominated scenario~\cite{Wands:1998yp, Finelli:2001sr}, and/or must involve additional degrees of freedom, as in the New Ekpyrotic scenario~\cite{Buchbinder:2007ad,Lehners:2007ac,Creminelli:2007aq}. 

The pseudo-conformal universe discussed in this paper exploits the latter loophole, introducing additional degrees of freedom as progenitors of density perturbations. 
Pseudo-conformal cosmology~\cite{Rubakov:2009np,Creminelli:2010ba,Hinterbichler:2011qk} is an alternative to inflation which postulates that the universe at very early times is cold, nearly static, and governed by an approximate conformal field
theory (CFT) on approximately Minkowski space. The conformal theory is invariant under the conformal algebra of $4$-dimensional Minkowksi space, namely $\mathfrak{so}(4,2)$. The central ingredient of the scenario
is that the dynamics allow for at least one scalar operator (of non-zero conformal weight) in the CFT to acquire a time-dependent expectation value which breaks the $\mathfrak{so}(4,2)$ algebra down to $\mathfrak{so}(4,1)$,
\be
\mathfrak{so}(4,2)\longrightarrow \mathfrak{so}(4,1)\,. 
\label{pseudobreaking}
\ee
Specifically, this symmetry breaking pattern follows from scalar operators $\phi_I$, $I = 1,\ldots, N$, of weight $\Delta_I\neq 0$ developing the time-dependent profile
\be 
\bar{\phi}_I(t) \sim \frac{1}{(-t)^{\Delta_I}}\,, 
\label{pseudoback}
\ee
where $-\infty < t < 0$, and $t=0$ signals the transition into a standard big bang phase.  Since the $\mathfrak{so}(4,1)$ unbroken symmetry algebra coincides with the algebra of isometries of de Sitter space, it is not surprising that certain fields (namely weight-0 fields) in the theory acquire scale invariant perturbations under very general conditions. These are entropy or isocurvature perturbations, which at some later stage must be converted to adiabatic perturbations through standard mechanisms~\cite{Lyth:2001nq, Kofman:2003nx, Dvali:2003em}.

Despite the appearance of $\mathfrak{so}(4,1)$, the scenario is {\it not} equivalent to inflation. The mechanism is intrinsically non-gravitational, and to a good approximation can be described on flat, Minkowski space-time.
In the presence of gravity, the Einstein-frame scale factor to which the CFT minimally couples is either very slowly contracting or expanding, corresponding to an equation of state $|w| \gg 1$. Such a phase of slow contraction
(or expansion) is well-known to make the universe increasingly flat, homogeneous and isotropic, akin to the smoothing mechanism in ekpyrotic cosmology~\cite{Khoury:2001wf}. 

As usual with spontaneous symmetry breaking, much of the relevant physics derives from the assumed symmetries, irrespective of the underlying microphysics.
The mechanism described above first appeared in explicit incarnations, namely the negative-$\phi^4$ model~\cite{Rubakov:2009np,Craps:2007ch} and Galilean
Genesis~\cite{Creminelli:2010ba}. As was pointed out in~\cite{Hinterbichler:2011qk}, however, the key phenomena encountered in these realizations follow from the symmetry
breaking pattern~(\ref{pseudobreaking}). In particular, it was shown that the quadratic action for the perturbations is completely fixed by the symmetries. 

In this paper, we systematically construct the most general low-energy effective action that linearly realizes $\mathfrak{so}(4,1)$ and non-linearly realizes $\mathfrak{so}(4,2)$.
 Techniques for the construction of non-linear realizations such as this were developed in the 60's for internal symmetries~\cite{Coleman:1969sm, Callan:1969sn} and later extended to space-time
symmetries~\cite{volkov}. After reviewing these techniques and giving a few motivational examples, we apply them to the symmetry breaking pattern of interest (\ref{pseudobreaking}). 
For broken space-time symmetries, as is the case here, it is well-known that the standard counting of Goldstone bosons fails~\cite{Low:2001bw}. Indeed, even
though~(\ref{pseudobreaking}) implies 5 broken symmetries, there is only one Goldstone, $\pi$. This is due to the fact that some Goldstone degrees of freedom are not actually independent and are related by so-called inverse Higgs constraints \cite{Ivanov:1975zq}, which reduce the number of dynamical fields. The coset construction allows
us to write down the most general effective action for $\pi$ and other ``matter" fields, including weight-0 fields, systematically in powers of derivatives.

As an example of the utility of the effective field theory formalism, we construct the most general lagrangian up to quartic order in derivatives for the Goldstone $\pi$. Although the intermediate steps are somewhat technical, the end result is surprisingly simple. The Goldstone action is given by
\begin{align}
\nonumber
S_\pi = \int {\rm d}^4x \sqrt{-\bar g_{\rm eff}}&\bigg[M_0^2\left(-\frac{1}{2}e^{2\pi}(\partial\pi)^2- H^2e^{2\pi}+ \frac{H^2}{2}e^{4\pi}\right)\\
&~~~~~+M_1\bigg((\bar\square\pi)^2+2\bar\square\pi(\partial\pi)^2+(\partial\pi)^4-4H^2(\partial\pi)^2\bigg)\\
\nonumber
&~~~~~+M_2\bigg((\partial\pi)^4+2\bar\square\pi(\partial\pi)^2-6H^2(\partial\pi)^2-12H^2\pi+3H^2e^{4\pi}\bigg)+\ldots\bigg]\,,
\end{align}
where $M_0,M_1, M_2,\ldots $ are arbitrary constant coefficients. All raised indices and covariant derivatives are with respect to the metric $\bar g^{\rm eff}_{\mu\nu}$, which is a metric on de Sitter space with Hubble parameter $H$.  As we will discuss, however, the metric $\bar g^{\rm eff}_{\mu\nu}$ is not the physical Einstein-frame metric which will minimally couple to matter in the later universe.  The physical metric is the flat metric $\eta_{\mu\nu}$, and $\bar g^{\rm eff}_{\mu\nu}$ should be thought of as parameterizing the background VEV's for the conformal fields.  We will see that this difference is what makes the conformal scenario different from standard inflation.

Similarly, it is possible to use the non-linear realization machinery to couple matter fields to the Goldstone field. As an example, we construct an invariant action up to fourth order in derivatives for a fiducial weight-0 spectator field, $\chi$,
\begin{align}
\nonumber
S_\chi = \int\rd^4x\sqrt{-\bar g_{\rm eff}}&\left[-\frac{\bar M^2_\chi}{2}e^{2\pi}(\partial\chi)^2+e^{4\pi}V(\chi)+a_1(\partial\chi)^4+a_2(\bar\square\chi)^2+\ldots\right.\\
&~~~~~+\bar M^2_0(\chi)\left(\frac{1}{2}e^{2\pi}(\partial\pi)^2+\frac{1}{2}e^{2\pi}\bar\square\pi-H^2e^{2\pi}+\frac{H^2}{2}e^{4\pi}\right)\\
\nonumber
&~~~~~+\left.\bar M_1(\chi)\bigg((\bar\square\pi)^2+2\bar\square\pi(\partial\pi)^2+(\partial\pi)^4-4H^2(\partial\pi)^2\bigg)+\ldots\right]~,
\end{align}
where the functions, $V(\chi), \bar M_0^2(\chi)\ldots$, are arbitrary polynomial functions of $\chi$. We assume that $V(0)=\rd V/\rd\chi \rvert_0=\rd \bar M_0^2/\rd\chi\rvert_0=0$ so that there are no tadpole contributions and $\pi=\chi=0$ is a consistent solution.

Similar to the effective field theory of inflation formalism \cite{Cheung:2007st}, the form of the various operators appearing in the theory is fixed by the symmetry breaking pattern while the various coefficients are of course model dependent. Nevertheless, this effective theory allows us to make general statements about the pseudo-conformal scenario. For example, we check that the 2-point function for $\chi$ is scale invariant for suitable choices of the couplings (it is this field that will feed into a scale-invariant spectrum for the adiabatic mode $\zeta$ after conversion).
The 2-point function for $\pi$, meanwhile, is strongly red tilted, corresponding to a de Sitter weight $-1$ field, 
consistent with earlier analyses~\cite{Rubakov:2009np, Creminelli:2010ba, Hinterbichler:2011qk}. The first non-trivial results arise at the 3-point level. The unbroken $\mathfrak{so}(4,1)$ symmetries act at late times as the
conformal group on $\mathbb{R}^3$, hence the 3-point function is completely fixed by conformal invariance~\cite{DiFrancesco:1997nk}, up to an overall normalization,
\be
 \lim_{t\rightarrow 0}\,\langle \varphi_1(\vec{x}_1,t) \varphi_2(\vec{x}_2,t)\varphi_3(\vec{x}_3,t)\rangle = \frac{C_{123}}{x_{12}^{\Delta_1 + \Delta_2-\Delta_3}x_{23}^{\Delta_2 + \Delta_3-\Delta_1}x_{13}^{\Delta_1 + \Delta_3-\Delta_2}}\,,
\ee
where $x_{ij} \equiv \lvert\vec{x}_i-\vec{x}_j\rvert$, the $\varphi$'s are any of the conformal fields, and the $\Delta$'s denote their conformal weights.  Using our effective action, we compute the 3-point functions $\langle\chi\chi\chi\rangle$ and $\langle\pi\chi\chi\rangle$ and find they are consistent
with $3d$ conformal invariance. Focusing only on the $\mathfrak{so}(4,1)$ symmetries, the form of correlation functions is identical to that of spectator fields in inflation~\cite{Antoniadis:2011ib,Creminelli:2011mw},
including gravitational waves~\cite{Maldacena:2011nz}, where the $\mathfrak{so}(4,1)$ isometries also act at late times as the conformal group on $\mathbb{R}^3$.

However, the correlation functions of pseudo-conformal cosmology also know about the full $\mathfrak{so}(4,2)$ symmetries and are therefore more constrained than their inflationary counterparts.
The 5 broken symmetries should result in Ward identities relating correlation functions with different numbers of fields, akin to the soft pion theorems of the chiral lagrangian for QCD.\footnote{Here we will again have `soft $\pi$' theorems, but the $\pi$ is of course a different field. For example, in the squeezed limit, the $\langle\pi\chi\chi\rangle$ correlator will be related to the two-point functions of $\pi$ and $\chi$.} The pseudo-conformal correlation functions may
therefore exhibit distinguishing relations that the smaller symmetry algebra of inflation cannot reproduce. We will describe the systematic derivation of these relations elsewhere, but as a glimmer of what such relations might entail, we show how the coefficients of particular interactions are fixed by symmetry and argue that these should lead to strict relations between the normalization of the four-point and three-point functions, for example.

After working out the effective action using standard non-linear realization techniques, we illustrate an alternative to the coset construction. This alternative technique is a straightforward generalization of the curvature invariant method employed in \cite{Nicolis:2008in} to construct the conformal galileons. Our desire is to construct actions which linearly realize the de Sitter group while also non-linearly realizing the full conformal group. A clear way to linearly realize the isometries of de Sitter is to construct a field theory on an effective, fictitious de Sitter space, $\bar{g}_{\mu\nu}^{\rm eff}$.  If we then add the conformal mode to the de Sitter metric and consider
\be
g_{\mu\nu}^{\rm eff} = e^{2\pi}\bar{g}_{\mu\nu}^{\rm eff}\,,
\label{confdsmetric}
\ee
then theories constructed from diffeomorphism invariants of this metric will have the symmetries of the conformal group of de Sitter space, which is the same group as the conformal group of flat space, {\it  i.e.}, the desired $\mathfrak{so}(4,2)$ symmetry..  Similarly, we may couple matter fields to the Goldstone field by using the geometric covariant derivative associated to the metric (\ref{confdsmetric}). We argue that this construction is entirely equivalent to the coset construction.

Again, it is worth stressing the difference between the effective de Sitter space which emerges in the effective action for perturbations, and the nearly Minkowskian physical space-time
$g_{\mu\nu}\simeq \eta_{\mu\nu}$ which describes the actual, Einstein-frame geometry, and which will minimally couple to matter in the later universe. As illustrated explicitly with the $\phi^4$ example in Sec.~\ref{Notinf}, the de Sitter metric is related to the Minkowskian, Einstein-frame metric
by $g_{\mu\nu}^{\rm eff} = \phi^2 g_{\mu\nu}\simeq \phi^2\eta_{\mu\nu}$. Of course one can perform a conformal transformation to work in terms of $g_{\mu\nu}^{\rm eff}$, whose background solution is de Sitter, but the resulting action is in Jordan frame and involves a strongly time-varying Newton's constant, showing that the scenario is distinctly different from inflation. Incidentally, the coexistence of a physical Minkowski geometry and a fictitious de Sitter metric also explains the necessity of having the full $\mathfrak{so}(4,2)$ conformal symmetries -- the conformal group is the smallest group which contains both a de Sitter and a Poincar\'e subgroup.

The paper is organized as follows. In Sec.~\ref{intro}, we review the pseudo-conformal mechanism in the simplest of cases, the negative quartic potential of~\cite{Rubakov:2009np, Hinterbichler:2011qk}. In an Appendix, we also give a novel six-dimensional viewpoint on the scenario in this simple case. In Sec.~\ref{Notinf}, we describe the cosmological dynamics and show explicitly that this scenario is not inflationary. In Sec.~\ref{coset}, we review the coset construction technique for non-linear realizations in both the case of internal and of space-time symmetry breaking. We give some instructional examples of the techniques, including the well-studied case where the conformal group is spontaneously broken to Poincar\'e. In Sec.~\ref{conftods} we apply the coset construction to the symmetry breaking pattern of principal interest, where the conformal group is broken to its de Sitter subgroup. In Sec.~\ref{curvinvariants} we corroborate the results of the coset construction by building the low-energy effective action using the curvature invariant technique. After constructing the actions for both the Goldstone and a weight-0 spectator field, we compute the 2-point functions for both in Sec.~\ref{lowenergyeffectiveaction}, and verify that the spectator field indeed has a scale-invariant spectrum of fluctuations. We also consider the 3-point functions $\langle\chi\chi\chi\rangle$ and $\langle\pi\chi\chi\rangle$. Finally, we summarize our results and discuss future directions in Sec.~\ref{conclusion}.
\section{Review of the Pseudo-Conformal Scenario}
\label{intro}
Before diving into the derivation of the effective action, it is worth reviewing the conformal scenario through its simplest realization: a conformal scalar field $\phi$ with {\it negative} $\phi^4$ potential.
The negative $\phi^4$ example was considered in the context of a holographic dual to an AdS$_5$ bouncing cosmology by~\cite{Craps:2007ch}, discussed in the
present context in a series of papers by Rubakov~\cite{Rubakov:2009np,Osipov:2010ee, Libanov:2010nk, Libanov:2010ci, Libanov:2011bk, Libanov:2011zy}, and further developed in~\cite{Hinterbichler:2011qk}. 

\subsection{Simplest Illustration of the Mechanism}
\label{quarticpot}

Consider the action
\be
S_\phi = \int\rd^4x\left[-\frac{1}{2}(\partial\phi)^2 + \frac{\lambda}{4}\phi^4\right]~,
\label{negquarticaction}
\ee
with ``wrong-sign" potential, $\lambda > 0$. The potential is unbounded from below, so we must imagine that higher-dimensional ({\it e.g.}, Planck-suppressed)
operators stabilize the field at large $\phi$~\cite{Hinterbichler:2011qk}. At the classical level, this theory is invariant under the 15 conformal transformations, under which $\phi$ is a field of weight $\Delta=1$, 
\be
\begin{array}{ll}
\delta_{P_\mu}\phi =-\partial_\mu\phi~,~~~~~~~~~ &
\delta_{J_{\mu\nu}}\phi = (x_\mu\partial_\nu-x_\nu\partial_\mu),\phi\\
\delta_D\phi = - (\Delta+ x^\mu\partial_\mu) \phi~,~~~~~~~~~ &
\delta_{K_\mu}\phi  = \left(-2\Delta x_\mu -2x_\mu x^\nu\partial_\nu +x^2\partial_\mu\right)\phi~.
\end{array}
\label{confgen}
\ee
These form the $\mathfrak{so}(4,2)$ algebra, as may be seen by repackaging the generators (\ref{confgen}) as
\be
\begin{array}{ll}
\delta_{J_{\mu\nu}} = \delta_{J_{\mu\nu}}~,~~~~~~~~~~~~~~~&
\delta_{J_{5\mu}} = \frac{1}{2}\left(\delta_{P_\mu}+\delta_{K_\mu}\right)~,\\
\delta_{J_{56}} = \delta_D~,~~~~~~~~~~~~~~~ &
\delta_{J_{6\mu}} = \frac{1}{2}\left(\delta_{P_\mu}-\delta_{K_\mu}\right)~,
\end{array}
\label{6dconfalg}
\ee
which then satisfy the $\mathfrak{so}(4,2)$ algebra
\be
\left[\delta_{J_{AB}}, \delta_{J_{CD}}\right] = \eta_{AC}\delta_{J_{BD}}-\eta_{BC}\delta_{J_{AD}}+\eta_{BD}\delta_{J_{AC}}-\eta_{AD}\delta_{J_{BC}}~,
\ee
where $\eta_{AB}={\rm diag}(\eta_{\mu\nu},1,-1)$. The equation of motion for the action (\ref{negquarticaction}), assuming a homogeneous field profile, is 
\be
\ddot\phi - \lambda\phi^3 = 0~,
\ee
which has the zero-energy solution
\be
\bar\phi(t) = \sqrt{\frac{2}{\lambda}}\frac{1}{(-t)}~.
\label{1otsoln}
\ee
This solution is a dynamical attractor~\cite{Hinterbichler:2011qk}, essentially because the growing mode solution for small perturbations $\delta\phi$ can
be absorbed at late times into a time shift of the background. The profile~(\ref{1otsoln}) spontaneously breaks the symmetry algebra of the action (\ref{negquarticaction}) to its $\mathfrak{so}(4,1)$ de Sitter subalgebra. Indeed, the subalgebra of conformal generators~(\ref{confgen}) that annihilate the background~(\ref{1otsoln}) is spanned by 
\be
\left\{\delta_{P_i},~\delta_D,~\delta_{J_{ij}},~\delta_{K_i}\right\}~.
\label{1otgens}
\ee
These can be packaged into the generators
\be
\delta_{J_{56}} = \delta_D,~~~~~~~~~~~~~~\delta_{J_{5i}} = \frac{1}{2}\left(\delta_{P_i}+\delta_{K_i}\right),~~~~~~~~~~~~~\delta_{J_{6i}}= \frac{1}{2}\left(\delta_{P_i}-\delta_{K_i}\right),
\ee
which have the commutation relations of the $\mathfrak{so}(4,1)$ algebra,
\be
\left[\delta_{J_{ab}}, \delta_{J_{cd}}\right] = \eta_{ac}\delta_{J_{bd}}-\eta_{bc}\delta_{J_{ad}}+\eta_{bd}\delta_{J_{ac}}-\eta_{ad}\delta_{J_{bc}}~,
\ee
where $\eta_{ab} = {\rm diag}\left(\delta_{ij}, 1, -1\right)$.

Now, let us consider coupling a weight-0 spectator, {\it i.e.}, a field $\chi$ which transforms under (\ref{confgen}) with $\Delta=0$ , to the rolling field $\phi$. In order for the action to be dilation invariant, the action for $\chi$ up to quadratic order (and second order in derivatives) must be of the form
\be
S_\chi = \int\rd^4x \left[-\frac{1}{2}\phi^2(\partial\chi)^2 -\frac{m^2_\chi}{2}\phi^4\chi^2+\kappa\phi\square\phi\chi^2\right]~.
\label{Schi}
\ee
In fact, this action is invariant under the full conformal group where $\chi$ transforms as a weight-0 field. When $\phi$ gets the profile (\ref{1otsoln}), we may think of the $\chi$ field as coupling via the effective metric
\be
g_{\mu\nu}^{\rm eff} = \bar\phi^2\eta_{\mu\nu}=  \frac{2}{\lambda t^2}\eta_{\mu\nu} ~,
\label{geff1}
\ee
which is the metric of de Sitter space in a flat slicing. Thus, the $\chi$ field feels as though it lives on de Sitter space. It is emphasized that this is {\it not} the physical metric -- everything takes place in flat Minkowski space. It should not be surprising in light of the fact that $\chi$ lives in an effective de Sitter space that it can acquire a scale-invariant spectrum of perturbations. Indeed, if $m_\chi$ and $\kappa$ are sufficiently small, in the long wavelength limit the power spectrum is \cite{Hinterbichler:2011qk} 
\be
\mathcal P_\chi = \frac{1}{2\pi^2}k^3\lvert\chi_k\rvert\simeq\frac{\lambda}{2(2\pi)^2}~,
\ee
which is indeed scale invariant. The key insight of \cite{Hinterbichler:2011qk} is that weight-0 fields acquiring a scale-invariant spectrum is a feature generic to the symmetry breaking pattern $\mathfrak{so}(4,2) \to \mathfrak{so}(4,1)$.

\section{Cosmological Dynamics -- Why This is Not Inflation}
\label{Notinf}

The pseudo-conformal scenario assumes that the CFT couples minimally to Einstein gravity,
\be
S = \int {\rm d}^4 x\sqrt{-g}\left( \frac{M_{\rm Pl}^2}{2} R  + {\cal L}_{\rm CFT}\left[g_{\mu\nu}\right]\right)\,.
\label{Seinframe}
\ee
Conformal invariance is thus (mildly) broken at the $1/M_{\rm Pl}$ level. (The above covariantization is consistent with
that assumed in the Galilean Genesis scenario \cite{Creminelli:2010ba}; in his $\phi^4$ example, Rubakov \cite{Rubakov:2009np} instead considers conformal coupling to gravity. Conformal coupling is also considered in \cite{Piao:2011bz,Piao:2011mq}, in a similar context.)  

The action~(\ref{Seinframe}) is cast in Einstein frame, where the Planck scale is constant and the metric will be nearly flat.  We first describe our cosmological background in this frame,
and then turn to a ``Jordan-frame" description in terms of the effective de Sitter geometry which the weight-0 spectators couple to.  Comparing the descriptions will make clear that the conformal scenario is dramatically different than inflation.

\subsection{Einstein-Frame Cosmology}

At sufficiently early times (to be made precise shortly), gravity is negligible, hence the solution~(\ref{pseudoback}) is approximately valid.
Since this background only depends on time and is invariant under dilatation, the pressure and energy density must both scale
as $1/t^4$. But  energy conservation implies $\rho \simeq {\rm const.}$ at zeroth order in $1/M_{\rm Pl}$, hence $\rho \simeq 0$.
Thus, the assumed symmetries completely fix the form of the energy density and pressure of the CFT,
\be
\rho_{\rm CFT} \simeq 0\;,\qquad P_{\rm CFT} \simeq \frac{\beta}{t^4}\,,
\label{rhoP}
\ee
up to a constant parameter $\beta$. For instance, for the quartic potential model discussed in Sec.~\ref{quarticpot}, $\beta = 2/\lambda > 0$ 
corresponding to positive pressure. In the Galilean Genesis scenario \cite{Creminelli:2010ba}, on the other hand, $\beta < 0$, and the CFT violates the
Null Energy Condition.

Integrating $M_{\rm Pl}^2\dot{H} = -(\rho_{\rm CFT} + P_{\rm CFT})/2$ gives the Hubble parameter
\be
H(t) \simeq \frac{\beta}{6t^3M_{\rm Pl}^2}\,,
\label{Hein}
\ee
which corresponds to a contracting or expanding universe depending on the sign of $\beta$. In particular, the universe is contracting in the quartic potential case
($\beta = 2/\lambda$), and expanding in the Galilean Genesis scenario $(\beta < 0)$. We can integrate once more to obtain the scale factor
\be
a(t) \simeq 1 - \frac{\beta}{12 t^2M_{\rm Pl}^2}\,.
\ee
This self-consistently shows that the universe is indeed nearly static at early times. Specifically, neglecting gravity is valid for $t\ll t_{\rm end}$, with
\be
t_{\rm end} \equiv -\frac{\sqrt{\beta}}{M_{\rm Pl}}\,.
\label{tend}
\ee
Note that in the $\phi^4$ example, for instance, this corresponds to $\phi(t_{\rm end}) \sim M_{\rm Pl}$, where one in any case expects $M_{\rm Pl}$ suppressed operators to
regulate the potential.

One last word about the cosmology in Einstein frame: the evolution~(\ref{Hein}) implies the CFT equation of state
\be
w_{\rm CFT} \simeq \frac{P_{\rm CFT}}{\rho_{\rm CFT}} = \frac{12}{\beta} t^2M_{\rm Pl}^2\,.
\ee
Over the range $-\infty < t < t_{\rm end}$, the equation of state decreases from $+\infty$ to a value of ${\cal O}(1)$. A contracting phase with $w \gg 1$ is characteristic of
ekpyrotic cosmologies. The key difference here compared to earlier ekpyrotic scenarios is that $w$ is rapidly decreasing in time, as opposed to
being nearly constant~\cite{Khoury:2001wf} or growing rapidly~\cite{Khoury:2009my, Khoury:2011ii, Joyce:2011ta}. A phase of contraction/expansion with $|w|\gg 1$ is well known to drive
the universe to be increasingly flat, homogeneous and isotropic~\cite{Gratton:2003pe}. Hence the background of interest is a dynamical attractor, even in the presence of gravity.

\subsection{Jordan-Frame de Sitter Description}

The above makes it clear that the cosmological evolution is non-inflationary, since in the Einstein frame, the scale factor is either slowly contracting or expanding. 
Nevertheless, since we have already mentioned that weight-0 spectator fields experience an effective de Sitter metric -- see~(\ref{geff1}) -- one may wonder
whether the scenario is secretly inflation when cast in terms of this other metric. To shed light on this issue, consider for concreteness a single time-evolving
scalar field $\phi$ of weight 1, as in the example of Section \ref{quarticpot}. As in~(\ref{Schi}), weight-0 fields are assumed to couple to an effective, ``Jordan-frame" metric\footnote{The effective metric $g_{\mu\nu}^{\rm eff}$ thus defined
carries units, but this is inconsequential to our arguments; alternatively, one could write $g_{\mu\nu}^{\rm eff} = (\phi^2/M^2)g_{\mu\nu}$ and carry the mass scale $M$ throughout the calculation.}
\be
g_{\mu\nu}^{\rm eff} = \phi^2 g_{\mu\nu}\,.
\label{geff2}
\ee
Let us see how the de Sitter background arises in Jordan frame. Upon the conformal transformation~(\ref{geff2}), the action~(\ref{Seinframe}) becomes
\be
S = \int {\rm d}^4 x\sqrt{-g_{\rm eff}}\left( \frac{M_{\rm Pl}^2}{2\phi^2} R_{\rm eff}  + \frac{3M_{\rm Pl}^2}{\phi^4}g_{\rm eff}^{\mu\nu}\partial_\mu\phi\partial_\nu\phi + \frac{1}{\phi^4} {\cal L}_{\rm CFT}\left[\phi^{-2}g_{\mu\nu}^{\rm eff}\right]\right) \,.
\label{Sjordan}
\ee
The Friedmann and scalar field equations that derive from~(\ref{Sjordan}) take the simple form
\bea
\nonumber
& & 3H_{\rm eff}^2 \simeq 6H_{\rm eff}\frac{\dot{\phi}}{\phi^2} - 3\frac{\dot{\phi}^2}{\phi^4}\,,\\
& & \frac{\ddot{\phi}}{\phi^3} + 3H_{\rm eff} \frac{\dot{\phi}}{\phi^2} - 3\frac{\dot{\phi}^2}{\phi^4} - \frac{R_{\rm eff}}{6} = - \frac{\beta}{4\phi^2M_{\rm Pl}^2t^4}\,, \label{jordanfrweq}
\eea
where $H_{\rm eff} = \phi^{-1} {\rm d}\ln a_{\rm eff}/{\rm d}t$ is the Jordan-frame Hubble parameter, and dots are time derivatives with respect to the time coordinate $t$ (we have not changed coordinates, only conformal frames).
We have used~(\ref{rhoP}) to substitute for the energy density and pressure of the CFT.

The $\beta$ term on the right hand side of the second equation of (\ref{jordanfrweq}) is suppressed by $1/M_{\rm Pl}$ and hence is negligible at sufficiently early times (specifically when $t\ll t_{\rm end}$ from (\ref{tend})). In this regime, the equations allow for a solution
$\phi \sim 1/t$ and $H_{\rm eff} = {\rm constant}$, consistent with the Einstein-frame analysis. Thus the effective geometry is indeed approximately de Sitter. But this is emphatically {\it not} inflation
in any usual sense. The de Sitter expansion results from the non-minimal coupling of $\phi$ to gravity in this Jordan frame. In particular, the effective Planck scale $M_{\rm Pl}^{\rm eff} \sim 1/\phi$
varies by order unity in a Hubble time.

\section{Phenomenological Lagrangians}
\label{coset}
We now turn to the systematic construction of actions realizing the symmetry breaking pattern (\ref{pseudobreaking}) of the conformal scenario.
Symmetry is a powerful tool in the study of physical phenomena. As illustrated most famously by the chiral lagrangian of the strong interactions~\cite{Weinberg:1968de}, much of the dynamics of physical systems follows solely from symmetry breaking patterns.  In two classic papers~\cite{Coleman:1969sm, Callan:1969sn},  Callan, Coleman, Wess and Zumino developed a general algorithm for constructing low-energy effective actions, the so-called coset construction. 
The original work dealt with internal symmetries, but was extended to the case of space-time symmetry breaking shortly thereafter by Volkov \cite{volkov}. Here we briefly review the coset construction in both the internal symmetry and space-time cases and present some simple examples. We will then apply the coset construction to the symmetry-breaking pattern of interest in Sec.~\ref{conftods}. Nice reviews of the coset construction are given by~\cite{xthschool, Zumino:1970tu}.

\subsection{Coset Construction for Internal Symmetries}
Consider a theory which is invariant under some continuous internal symmetry group G, which is spontaneously broken to some continuous subgroup H. The Goldstone fields then parameterize the coset space ${\rm G}/{\rm H}$. Following \cite{Coleman:1969sm, Callan:1969sn, volkov}, we want to write down the most general H-invariant lagrangian which non-linearly realizes the G symmetry. 

The Lie algebra $\mathfrak g$ of G admits an orthogonal decomposition
\be
\mathfrak{g} = \mathfrak{h}\oplus\mathfrak{a}~,
\ee
where $\mathfrak h$ is the Lie algebra of the preserved group H, and $\mathfrak a$ is its orthogonal complement.\footnote{This decomposition is orthogonal with respect to the inner product given by the Killing form. When dealing with matrix realizations of the algebra, this is usually just the trace of the product of two matrices.} We denote bases of these subspaces as $V_i \in \mathfrak h$ and $A_a \in\mathfrak a$. A convenient parametrization of the Goldstone fields is given by 
\be
g(x) = e^{\xi(x)\cdot A}~,
\label{hsection}
\ee
where $\xi(x)\cdot A \equiv \xi^a(x) A_a$. The $\xi^a(x)$'s denote real scalar fields (the Goldstones) which are 
allowed to depend on space-time coordinates. The fields transform under a left action by $\bar g\in {\rm G}$ as
\be
\bar gg = \bar ge^{\xi(x)\cdot A} = e^{\xi'(x,\bar g)\cdot A}h\left(\xi,\bar g\right).
\label{gtrans}
\ee
 The appearance of $h(\xi, \bar g) \in {\rm H}$ preserves the parametrization (\ref{hsection}).  On G, there is a distinguished left-invariant Lie algebra-valued 1-form, the Maurer--Cartan form,
\be
\omega = g^{-1}\rd g = \omega_V^i V_i+\omega_A^aA_a~,
\ee
where we have expanded $\omega$  in the basis of the Lie algebra $\mathfrak g$. Now, $\omega$ is invariant under left G-transformations, but shifts under the local right H-transformations in (\ref{gtrans}). Its components therefore shift non-linearly under the transformation (\ref{gtrans}) as\footnote{We are assuming that the Lie algebra satisfies $[\mathfrak{h},\mathfrak{a}]\sim \mathfrak{a}$.}
\begin{align}
\nonumber
\omega_A^aA_a &\longmapsto h(\xi, \bar g)\left[\omega_A^aA_a\right] h^{-1}(\xi,\bar g)\\
\omega_V^i V_i  &\longmapsto h(\xi, \bar g)\left[\omega_V^i V_i + \rd\right]h^{-1}(\xi, \bar g)~.
\end{align}
We see that $\omega_A^a$ transforms covariantly and so provides an ingredient for constructing invariant lagrangians.  Any lagrangian that is constructed to be H-invariant will automatically non-linearly realize G.  We can think of $\omega_A^a$ as the covariant derivative for the Goldstone fields $\xi^a$,
\be
\rd x^\mu\mathcal D_\mu\xi^a(x) = \omega_A^a~.
\ee
On the other hand, $\omega_V^i$ transforms as a gauge connection.  From these, we can construct higher covariant derivatives for the Goldstone fields \cite{Coleman:1969sm, Callan:1969sn, xthschool}
as well as for other ``matter" fields, $\psi$, transforming in some representation $D$ of H:
\be
\rd x^\mu\bar{\mathcal D}_\mu\psi(x)= \rd\psi(x) + \omega_V^i D(V_i)\psi(x)\,.
\ee
From these ingredients, we can construct the most general lagrangian which is invariant under H and non-linearly realizes G.  In summary, the building blocks are the following objects,
\be
\mathcal D_\mu\xi^a~,~~~~~~~~~\bar{\mathcal D}_\mu~,~~~~~~~~~\psi~,~~~~~~~~~\eta_{\mu\nu}~,
\ee
along with any invariant tensors of the group ${\rm H}$ (and possibly the epsilon tensor, if one does not care about parity). An invariant lagrangian is then built out of terms which are both Lorentz-covariant and have fully contracted internal H indices.
\subsection{Spontaneously Broken Space-time Symmetries}
\label{spacetimecoset}
The coset construction in the case of spontaneously broken space-time symmetries is similar, but there are various subtleties. Here we give a brief review following~\cite{xthschool}. 

Consider a symmetry group ${\rm G}$ which contains some unbroken generators of space-time translations, $P_\mu$, Lorentz transformations $J_{\mu\nu}$, and some unbroken symmetry group ${\rm H}$ generated by $V_i$. Furthermore, we assume there are some broken generators, $Z_a$. In writing the coset element, we treat the unbroken $P_\mu$'s on essentially the same footing as the broken symmetry generators $Z_a$'s, since
the coordinates $x^\mu$ transform non-linearly under translations.  Hence we parameterize the coset ${\rm G}/{\rm H}$ by
\be\label{spacetimeparam}
g = e^{x\cdot P}e^{\xi(x)\cdot Z}~.
\ee
A left ${\rm G}$-transformation acts as \cite{xthschool, volkov}
\be
\bar ge^{x\cdot P}e^{\xi\cdot Z} = e^{x'\cdot P}e^{\xi'(x')\cdot Z}~h(\xi(x),\bar g)~,
\label{trans}
\ee
where
\be
h(\xi(x),\bar g) = e^{u(\xi, \bar g)\cdot V}e^{w(\xi, \bar g)\cdot J}
\label{hgbar}
\ee
is an element of the unbroken group H. 

Thus far the discussion parallels the internal symmetry case, except for the way we have dealt with space-time translations. In much the same way as for internal symmetries, if we restrict the ${\rm G}$-transformation to the unbroken group ${\rm H}$, we find that the symmetries are linearly realized. It is also easy to see that under a translation, the space-time coordinates transform inhomogeneously $x^\mu\to x^\mu + c^\mu$. It is for this reason that we choose to treat the $x^\mu$ in the same way as the Goldstone fields above. Another way to see that this is useful is to recall that space-time may be viewed as the coset Poincar\'e$/$Lorentz~\cite{Low:2001bw}.

As before, the appropriate object to consider is the Maurer--Cartan $1$--form,
\be
\omega = g^{-1}{\rm d}g = \omega_P^\mu P_\mu+\omega_Z^aZ_a+ \omega_V^iV_i+{1\over 2}\omega_J^{\mu\nu}J_{\mu\nu}~.
\ee
As in the internal symmetry case, the Maurer--Cartan form is left-invariant under global G-transformations. It is, however, not invariant under local action by H on the right. The transformation rules for the forms are~\cite{xthschool}
\begin{align}
\nonumber
\omega_P^\mu P_\mu&\longmapsto h(\xi, \bar g)\left[\omega_P^\mu P_\mu\right]h^{-1}(\xi, \bar g)\,,\\
\omega_Z^aZ_a&\longmapsto h(\xi, \bar g)\left[\omega_Z^aZ_a\right] h^{-1}(\xi, \bar g)\,,\\
\nonumber
\omega_V^iV_i+{1\over 2}\omega_J^{\mu\nu}J_{\mu\nu}&\longmapsto h(\xi, \bar g)\left[\omega_V^iV_i+{1\over 2}\omega_J^{\mu\nu}J_{\mu\nu}+{\rm d}\right]h^{-1}(\xi, \bar g)~,
\end{align}
where $h(\xi, \bar g)$ is as in~(\ref{hgbar}). From this, we can deduce the form of the covariant derivatives for the Goldstone modes. 

An additional subtlety in the space-time case is that the 1-forms ${\rm d}x^\mu$ no longer have simple transformation rules under broken transformations. As a result, the appropriate basis 1-forms to use are $\omega_P^\mu$, which transform in the correct way \cite{volkov, Low:2001bw, xthschool}. One can think of $\omega_P^\mu$ as giving a vielbein by writing $\omega_P^\alpha = e^{\ \alpha}_\mu {\rm d}x^\mu$. Using this, we can construct the covariant derivative of Goldstone fields \cite{xthschool, Bellucci:2002ji}
\be
\omega_P^\mu \mathcal D_\mu\xi^a = \omega_Z^a~,
\ee
and for matter fields of any ${\rm H}$ representation and Lorentz representation
\be
\omega_P^\mu \bar{\mathcal D}_\mu\psi = {\rm d}\psi+\omega_V^iD(V_i)\psi+{1\over 2}\omega_J^{\mu\nu}D(J_{\mu\nu})\psi~.
\ee
Note that the left hand side is exactly the same as in the internal symmetry case except for the fact that we have $\omega_P^\mu$ multiplying $\mathcal D_\mu$ instead of just $\rd x^\mu$. As before, the building blocks for constructing actions are
\be
\mathcal D_\mu\xi^b~,~~~~~~~~~~\bar{\mathcal D}_\mu~,~~~~~~~~~~\psi~,~~~~~~~~~~e_\mu^{\ \alpha}~.
\ee
By construction, any scalar obtained by contracting these building blocks will non-linearly realize ${\rm G}$ and linearly realize ${\rm H}$.
The appropriate integration measure for the action is given by the determinant of the vielbein \cite{xthschool}
\be
-\frac{1}{4!}\epsilon_{\alpha\beta\gamma\delta}\omega_P^\alpha\wedge\omega_P^\beta\wedge\omega_P^\gamma\wedge\omega_P^\delta = {\rm d}^4x~{\rm det}~e~.
\ee
For low derivative-order actions, a computationally more straightforward tack is to build an invariant lagrangian by directly combining together the forms appearing in the Maurer--Cartan form with the wedge product. We will take advantage of this technique whenever it is more convenient.
\subsection{Inverse Higgs Constraints}
The coset construction in the space-time symmetry case is complicated by the fact that not all of the Goldstone fields are independent physical degrees of freedom. Although in the case of internal symmetries there is a Goldstone mode for each broken symmetry generator, this need not be true for broken space-time symmetries. In general we may be able to relate one Goldstone field to derivatives of another one. This is the so-called inverse Higgs effect~\cite{Ivanov:1975zq}. An intuitive  illustration of this phenomenon is the example of a co-dimension one domain wall in three spatial dimensions~\cite{Low:2001bw}. The domain wall clearly breaks both the translation and two rotations transverse to it, so na\"ively there should be a total of three Goldstone modes. However, if one performs an infinitesimal, spatially-varying translation of the wall, which is physically what a Goldstone boson represents, it is clear that this looks locally the same as an infinitesimal rotation of the wall.  The two modes are thus degenerate, and there is only one independent mode. This is a simple example of the general phenomenon when space-time symmetries are broken.

In practice, implementing the inverse Higgs constraint is not complicated~\cite{Ivanov:1975zq}:
if the commutator of an unbroken translation generator $P_\mu$ and a broken symmetry generator $A$ has a component along a different broken generator $B$, 
\be
\left[ P_\mu, A\right] \sim B+\ldots~,
\ee
then we may eliminate the Goldstone mode associated to $A$. This is done by setting the element of the Maurer--Cartan form associated with $B$ to zero, giving a relation between the Goldstone modes of $A$ and $B$:
\be
\omega_B = 0.
\ee
Note that this never happens in the internal symmetry case because the broken internal symmetry generators commute with the action of all space-time symmetry generators. The constraint is invariant under the action of the non-linearly realized group ${\rm G}$, and may therefore be consistently implemented in actions -- lagrangians which are invariant to start with will continue to be invariant after implementation of this constraint. In many cases of physical interest, this constraint is in fact equivalent to eliminating the unphysical Goldstones via their equation of motion~\cite{McArthur:2010zm}, so that the lagrangians before and after implementing the constraint are dynamically equivalent.

\subsection{Warm-Up: Breaking Conformal Symmetry to Poincar\'e}
As an example of the non-linear realization technique as applied to broken space-time symmetries, it is illuminating to consider the breaking pattern where the conformal algebra is spontaneously broken to its Poincar\'e subalgebra,
\be
\mathfrak{so}(4,2)\longrightarrow \mathfrak{iso}(3,1)~.
\ee
This symmetry breaking pattern was considered originally in \cite{volkov} and also extensively in \cite{Salam:1970qk, Isham:1970xz, Isham:1970gz}; see \cite{Low:2001bw,  Ivanov:1975zq,xthschool, Bellucci:2002ji, McArthur:2010zm} for nice discussions. Not only is this symmetry breaking pattern interesting in its own right, it will provide a nontrivial check in the case of breaking to the de Sitter subalgebra, as the construction must reproduce the results of this section in appropriate limits.

The conformal algebra in the standard basis is given by
\be
\begin{array}{ll}
\left[D, P_\mu\right] = -P_\mu \,,&
\left[D, K_\mu\right] = K_\mu\,, \\
\left[J_{\mu\nu}, K_\sigma\right] = \eta_{\mu\sigma}K_\nu-\eta_{\nu\sigma}K_\mu\,, &
\left[J_{\mu\nu},P_\sigma\right] = \eta_{\mu\sigma}P_\nu-\eta_{\nu\sigma}P_\mu\,, \\
\left[K_\mu,P_\nu\right] = 2J_{\mu\nu}-2\eta_{\mu\nu}D\,, &
\left[J_{\mu\nu}, J_{\rho\sigma}\right] = \eta_{\mu\rho}J_{\nu\sigma}-\eta_{\nu\rho}J_{\mu\sigma}+\eta_{\nu\sigma}J_{\mu\rho}-\eta_{\mu\sigma}J_{\nu\rho}\,.
\end{array}
\ee
The unbroken subalgebra is the Poincar\'e algebra, and we parameterize the coset space ${\rm G}/{\rm H}$ by\footnote{This differs slightly from the form (\ref{spacetimeparam}) since we use the product of two exponentials for broken generators, but this is an equally good and more convenient parametrization for our purposes.}
\be
g = e^{x^\mu P_\mu}e^{\pi D}e^{\xi^\mu K_\mu}~.
\ee
The Maurer--Cartan 1-form is given by~\cite{volkov, Bellucci:2002ji, McArthur:2010zm}
\be
\omega = g^{-1}{\rm d}g = \omega_P^\mu P_\mu + \omega_D D+\omega_K^\mu K_\mu + {1\over 2}\omega_J^{\mu\nu}J_{\mu\nu}~,
\ee
with components
\bea
\nonumber
\omega_P^\mu &=& e^\pi {\rm d}x^\mu\,,\\
\nonumber
\omega_D &=& {\rm d}\pi+2e^\pi\xi_\mu{\rm d}x^\mu\,,\\
\nonumber
\omega_K^\mu &=& {\rm d}\xi^\mu+\xi^\mu{\rm d}\pi+e^\pi\left(2\xi^\mu\xi_\nu{\rm d}x^\nu-\xi^2{\rm d}x^\mu\right)\,,\\
\omega_J^{\mu\nu} &=& -4e^\pi\left(\xi^\mu{\rm d} x^\nu-\xi^\nu{\rm d} x^\mu\right)~.
\eea
Here space-time indices have been raised and lowered with $\eta_{\mu\nu}$. There are at face value five Goldstone fields ($\pi$ and $\xi^\mu$), corresponding to the 5 broken symmetries, but four of them are redundant.
Indeed, in the commutator $\left[K_\mu,P_\nu\right] = 2J_{\mu\nu}-2\eta_{\mu\nu}D$, the dilation generator appears on the right, which implies that we can eliminate the Goldstone fields associated with the broken special conformal transformations by implementing the inverse Higgs constraint,
\be
\omega_D = 0~,
\ee
giving a relation between the Goldstones
\be
\xi_\mu = -\frac{1}{2}e^{-\pi}\partial_\mu\pi~.
\label{confinversehiggs}
\ee
Hence there is only one independent Goldstone boson -- the dilaton $\pi$.

We are free to substitute~(\ref{confinversehiggs}) back into the Maurer--Cartan form at will. In this way, it is convenient to rewrite the relevant components as \cite{Bellucci:2002ji}
\be
\begin{array}{l}
\omega_P^\mu = e^\pi {\rm d}x^\mu = e^\pi \delta^\mu_\nu \rd x^\nu\,;\\
\omega_K^\mu = {\rm d}\xi^\mu-e^\pi\xi^2{\rm d}x^\mu\,;\\
\omega_J^{\mu\nu} = -4e^\pi\left(\xi^\mu{\rm d} x^\nu-\xi^\nu{\rm d} x^\mu\right)~.
\end{array}
\label{confmcforms}
\ee
Here we have only substituted the inverse Higgs constraint in places where it leads to algebraic simplifications, but all appearances of $\xi^\mu$ should be taken to be implicitly in terms of $\pi$, through (\ref{confinversehiggs}). Note that the vielbein can be readily extracted from $\omega_P^\mu$ to obtain $e_\nu^{\ \mu} = e^\pi\delta_\nu^\mu$. This yields the invariant metric 
\be
g_{\mu\nu} = e_\mu^{\ \rho} e_\nu^{\ \sigma}\eta_{\rho\sigma}= e^{2\pi}\eta_{\mu\nu}.
\ee
Meanwhile, the invariant measure is
\be
-\frac{1}{4!}\epsilon_{\mu\nu\rho\sigma} \omega_P^{\mu}\wedge\omega_P^{\nu}\wedge\omega_P^{\rho}\wedge\omega_P^{\sigma} = \rd^4x~e^{4\pi} = \rd^4x\sqrt{-g} ~.
\label{measure}
\ee
In order to construct the lagrangian for the dilaton $\pi$, we note that the covariant derivative associated to $\xi^\mu$ is given by the expression
\be
\omega_K^\mu = \omega^\nu_P\mathcal D_\nu\xi^\mu~.
\ee
This can be solved for the covariant derivative using the forms~(\ref{confmcforms}):
\be
\mathcal D_\nu\xi_\mu = e^{\pi}\partial_\nu\xi_\mu-e^{2\pi}\eta_{\alpha\beta}\xi^\alpha\xi^\beta\eta_{\mu\nu}=\frac{1}{2}\partial_\nu\pi\partial_\mu\pi-\frac{1}{2}\partial_\nu\partial_\mu\pi-\frac{1}{4}\eta^{\alpha\beta}\partial_\alpha\pi\partial_\beta\pi\eta_{\mu\nu}~.
\label{dxi}
\ee
Note that to construct invariant lagrangians from $\mathcal D_\nu\xi_\mu$, indices should be contracted with $g_{\mu\nu}$. This implicitly gives the covariant derivative of the physical Goldstone field $\pi$ after eliminating $\xi^\mu$ through the inverse Higgs constraint (\ref{confinversehiggs}) in the second equality. The covariant derivative of a matter field $\psi$ is similarly given by
\be
\omega_P^\mu\bar{\mathcal D}_\mu\psi = {\rm d}\psi + {1\over 2}\omega_J^{\mu\nu}D(J_{\mu\nu})\psi~.
\ee
The matter covariant derivative allows us to take higher derivatives of the object $\mathcal D_\nu\xi^\mu$. In this case, it is just the geometric covariant derivative $\nabla_\mu$ associated to the metric $g_{\mu\nu}$.

We are now in a position to construct the action for $\pi$. The building blocks are
\be
\mathcal D_\nu \xi_\mu~,~~~~~~~~~~~~g_{\mu\nu} = e^{2\pi}\eta_{\mu\nu}~,~~~~~~~~~~~~\bar{\mathcal D}_\mu = \nabla_\mu.
\label{confpoincareblocks}
\ee
Invariant actions consist of Lorentz contractions of these objects, multiplied by the invariant measure~(\ref{measure}). The simplest term has no derivatives
\be
S_0 =  M_{\rm v}^4\int {\rm d}^4x~\sqrt{-g} = M_{\rm v}^4\int {\rm d}^4x~e^{4\pi}\,.
\label{S0Poincare}
\ee
The next simplest term is the kinetic term\footnote{In Sec.~\ref{spacetimecoset} it was mentioned briefly that it is possible to construct actions directly from the Maurer--Cartan form. Here we can make this statement more explicit by considering the 4-form
\be
\nonumber
S_1 = -\frac{M_0^2}{3!}\int\epsilon_{\mu\nu\rho\sigma} \omega_K^{\mu}\wedge\omega_P^{\nu}\wedge\omega_P^{\rho}\wedge\omega_P^{\sigma}~,
\ee
which gives the kinetic term (\ref{S1Poincare}). In this way, we see that actions may be built by directly combining the invariant 1-forms using the Lorentz-invariant tensors $\eta_{\mu\nu}$ and $\epsilon_{\mu\nu\rho\sigma}$.}
\be
S_1 = M_0^2\int {\rm d}^4x~e^{4\pi}g^{\mu\nu}\mathcal D_{\mu}\xi_\nu = M_0^2\int {\rm d}^4x~\frac{1}{2}e^{2\pi}\left(\partial\pi\right)^2~,
\label{S1Poincare}
\ee
where the last step follows from integration by parts. This is the well-known expression for the kinetic term that non-linearly realizes conformal symmetry, first derived in~\cite{volkov}.

At the four-derivative level, we have\footnote{As a wedge product, this action corresponds to the sum of 4-forms
\be
\nonumber
S_2 = \int \left(\frac{1}{2}\epsilon_{\mu\nu\rho\sigma}\omega^\mu_K\wedge\omega^\nu_K\wedge\omega_P^\rho\wedge\omega_P^\sigma
+\frac{1}{3!}\eta_{\mu\nu}\omega_K^\mu\wedge\star_4\omega_K^\nu\right)\,,
\ee
where $\star_4$ is the Hodge dual with respect to the conformal metric, $\star_4\omega_K^\alpha = \frac{1}{3!}\epsilon_{\mu_0\mu_1\mu_2\mu_3}\mathcal D^{\mu_0}\xi^\alpha\omega_P^{\mu_1}\wedge\omega_P^{\mu_2}\wedge\omega_P^{\mu_3}$.}
\be
S_2 = \int {\rm d}^4x~e^{4\pi}\left(\mathcal D_{\mu}\xi^\mu\right)^2 = \frac{1}{4}\int\rd^4x\Big[(\square\pi)^2+2\square\pi(\partial\pi)^2+(\partial\pi)^4\Big]~.
\ee
It can be checked that this combination is is indeed conformally invariant. 
As we will show explicitly in Section \ref{conftods}, the other four-derivative term $\left(\mathcal D_\mu\xi_\nu\right)^2$ is not linearly independent in $d=4$, however one can construct a linearly independent term by taking a suitable limit as $d\to4$ (this same subtlety arises in a different guise in \cite{Nicolis:2008in}).
Terms of higher order in derivatives can be constructed by following the same pattern we have outlined above, building Lorentz scalars from the objects (\ref{confpoincareblocks}).

An alternative viewpoint on the coset construction, which will be explored more fully in Sec.~\ref{curvinvariants}, is based on the effective conformal metric
\be
g_{\mu\nu} = e^{2\pi}\eta_{\mu\nu}~.
\ee
Note that the Ricci tensor associated to this conformal metric,
\be
R_{\mu\nu} = 2\partial_\mu\pi\partial_\nu\pi-2\partial_\mu\partial_\nu\pi-\square\pi\eta_{\mu\nu}-2(\partial\pi)^2\eta_{\mu\nu}~,
\ee
can be expressed in terms of the covariant derivative $\mathcal D_\mu\xi^\nu$ after lowering an index
\be
4\mathcal D_\mu\xi_\nu+2\mathcal D_\alpha\xi^\alpha g_{\mu\nu} = R_{\mu\nu}~.
\ee
Additionally, $\mathcal D_\alpha\xi^\alpha \sim R$, the Ricci scalar for the conformal metric. We therefore see that the invariant action constructed by the coset method corresponds to all possible diffeomorphism scalars constructed from the metric $g_{\mu\nu} = e^{2\pi}\eta_{\mu\nu}$, its curvature tensors and its covariant derivative.

\section{Breaking Conformal to de Sitter}
\label{conftods}
We now turn to the case of principal interest -- spontaneously breaking the conformal algebra to its de Sitter subalgebra
\be
\mathfrak{so}(4,2) \longrightarrow \mathfrak{so}(4,1)~.
\label{symbreakbis}
\ee
To our knowledge, the coset construction for this symmetry breaking pattern has not appeared
previously in the literature. (The case of breaking conformal to the {\it Anti}-de Sitter algebra $\mathfrak{so}(3,2)$ was considered in \cite{Clark:2005ht}.) To this end, it is convenient to
parameterize the conformal algebra by the generators $J_{\mu\nu}$, $K_\mu$, $D$ and
\be
\hat P_\mu \equiv P_\mu + \frac{1}{4}H^2K_\mu\,,
\ee
where the dimensionful parameter $H$ will turn out to be the Hubble constant for the effective de Sitter metric. In this basis, the algebra takes the form
\be
\begin{array}{ll}
\left[\hat P_\mu, \hat P_\nu\right] = H^2J_{\mu\nu}, &
\left[D, \hat P_\mu\right] = -\hat P_\mu+\frac{1}{2}H^2K_\mu,\\
\left[D, K_\mu\right] = K_\mu, &
\left[\hat P_\mu, K_\nu\right] = 2\eta_{\mu\nu}D+2J_{\mu\nu},\\
\left[J_{\mu\nu}, K_\rho\right] = \eta_{\mu\rho}K_\nu - \eta_{\nu\rho}K_\mu ,&
\left[J_{\mu\nu}, \hat P_\rho\right] = \eta_{\mu\rho}\hat P_\nu - \eta_{\nu\rho}\hat P_\mu,\\
\left[ J_{\mu\nu}, J_{\sigma\rho}\right] = \eta_{\mu\sigma}J_{\nu\rho}-\eta_{\nu\sigma}J_{\mu\rho}+\eta_{\nu\rho}J_{\mu\sigma}-\eta_{\mu\rho}J_{\nu\sigma}.
\end{array}
\ee
This parameterization of the conformal algebra appears also in \cite{Bellucci:2002ji} in the context of breaking the conformal algebra to Poincar\'e.
The advantage of working with $\hat P_\mu$ rather than the $P_\mu$ is that the set $\{\hat P_\mu, J_{\nu\rho}\}$ generates an $\mathfrak{so}(4,1)$ subalgebra.\footnote{Although this is not our main focus, one might also be interested in breaking the conformal algebra to its Anti-de Sitter subalgebra $\mathfrak{so}(3,2)$. This breaking pattern follows straightforwardly by defining $\bar P_\mu \equiv P_\mu - \frac{1}{4}H^2K_\mu$. Then, the set of generators $\{\bar P_\mu, J_{\nu\rho}\}$ generates an $\mathfrak{so}(3,2)$ subalgebra of $\mathfrak{so}(4,2)$. This symmetry breaking pattern was considered in \cite{Clark:2005ht}, using a different parameterization of the algebra. In order to obtain actions equivalent to theirs (but algebraically simpler), one can analytically continue $H^2 \to -H^2$ in the following sections.} This can be made manifest
by adding a fifth index and writing $J_{5\mu} \equiv \hat P_\mu$, in terms of which the commutation relations of $\{\hat P_\mu, J_{\nu\rho}\}$ take the $\mathfrak{so}(4,1)$ form,
\be
\left[J_{ab}, J_{cd}\right] = \eta_{ac}J_{bd}-\eta_{bc}J_{ad}+\eta_{bd}J_{ac}-\eta_{ad}J_{bc}~,
\ee
where $\eta_{ab} = {\rm diag}\left( -1, 1, 1, 1, 1\right)$ is the metric of 4+1 dimensional Minkowski space.
\subsection{Constructing the Effective Action}
\label{dsgoldlag}
Since the broken symmetries correspond to $D$ and $K_\mu$ in this basis, we parameterize the group coset by
\be
g = e^{y\cdot \hat P}e^{\pi D}e^{\xi\cdot K}~,
\label{dscosetsection}
\ee
where the inner product is taken with respect to the vielbein metric $\eta_{mn}$. As we will see shortly, the space-time coordinates $y^\mu$ corresponding to $\hat P_\mu$ parametrize a particular coordinate system on de Sitter space.
At the end of the day, however, it will be possible to express all of our results in a coordinate-independent way.

We can pull back the Maurer--Cartan form on the conformal group by this local section and expand it in components,
\begin{align}
\nonumber
\omega_{\hat P}^m &= e^\pi \bar e^m_\mu\rd y^\mu\;,\\
\nonumber
\omega_D &= \rd\pi + 2e^\pi\xi_m \bar e^m_\mu\rd y^\mu\;,\\
\nonumber
\omega_K^m &= \rd\xi^m-\omega^{mn}_{\rm spin}\xi_n+2e^\pi\xi_n\xi^m \bar e_\mu^n\rd y^\mu - e^\pi\xi^2 \bar e_\mu^m\rd y^\mu - \frac{H^2}{2}\sinh\pi \bar e_\mu^m\rd y^\mu+\xi^m\rd\pi\;,\\
{1\over 2}\omega_J^{mn} &= e^\pi\rd y^\mu\left(\xi^n\bar e^m_\mu - \xi^m \bar e^n_\mu\right)+ \omega^{mn}_{\rm spin}~.
\label{dsmcform}
\end{align}
Here, the vielbein is given by $e_\mu^m = e^\pi \bar e_\mu^m$ where,
\be
\bar e^m_\mu(y) = \left(\delta_\mu^m-\frac{y_\mu y^m}{y^2}\right) \frac{\sin\sqrt{H^2y^2}}{\sqrt{H^2y^2}}+\frac{y_\mu y^m}{y^2}~,
\ee
and the spin connection on de Sitter is given by
\be
\omega_{\rm spin}^{mn}(y) =\rd y^\mu\omega_\mu^{mn} = \left(\cos\sqrt{H^2y^2}-1\right)\left[\frac{y^n\rd y^m-y^m\rd y^n}{y^2}\right]~.
\ee
Although this is by no means obvious, these represent a vielbein and spin connection for de Sitter space. To see this explicitly, consider 
the coordinate transformation~\cite{Clark:2005ht}
\be
y^\mu = x^\mu \sqrt{\frac{4}{H^2x^2}} \arctan\sqrt{\frac{H^2x^2}{4}}~.
\ee
This brings the vielbein into diagonal form
\be
\bar e_\mu^m(x) = \left(\frac{1}{1+\frac{1}{4}H^2x^2}\right)\delta_\mu^m~,
\ee
corresponding to the better-known coordinitization of de Sitter with metric
\be
\bar g_{\mu\nu}^{\rm eff} = \left(\frac{1}{1+\frac{1}{4}H^2x^2}\right)^2\eta_{\mu\nu}~.
\ee
This makes it clear that the $y^\mu$ coordinates are in fact coordinates on de Sitter space, as claimed earlier. With this knowledge at hand, 
we can leave the coordinates arbitrary and consider a general de Sitter metric
\be
\bar g_{\mu\nu}^{\rm eff} = \bar e_\mu^m \bar e_\nu^n\eta_{mn}~,
\ee
allowing us to write everything in terms of space-time indices.

As we are now used to, there is an inverse Higgs constraint to be implemented which will give a relation between Goldstone fields. The commutator 
\be
[\hat P_\mu, K_\nu] = 2\eta_{\mu\nu}D+2J_{\mu\nu}
\ee
implies that the Goldstone fields $\xi^\mu$ associated to the $K_\mu$'s can be removed in favor of $\pi$. This is implemented by setting $\omega_D = 0$, which gives the relation\footnote{Although the form of the relation is the same as in the case where the conformal group is broken to Poincar\'e, here the space-time indices should be understood as being raised and lowered with a de Sitter metric instead of the flat metric.}
\be
\xi_\mu = -\frac{1}{2}e^{-\pi}\partial_\mu\pi\,.
\label{dSHiggs}
\ee
The expression~(\ref{dsmcform}) for the Maurer--Cartan form thus simplifies,
\begin{align}
\nonumber
\omega_{\hat P}^\mu &= e^\pi \rd y^\mu\;,\\
\nonumber
\omega_D &= \rd\pi + 2e^\pi\xi_\mu\rd y^\mu\;,\\
\nonumber
\omega_K^\mu &=\rd y^\nu\bar\nabla_\nu\xi^\mu - e^\pi\xi^2\rd y^\mu -\frac{H^2}{2}\sinh\pi\rd y^\mu\;, \\
{1\over 2}\omega_{\mu~J}^{ab} &= e^\pi\left(\xi^be^a_\mu - \xi^a e^b_\mu\right)+ \omega^{ab}_{\mu~{\rm spin}}~,
\label{dsmcform1}
\end{align}
where the contraction $\xi^2 = \bar{g}_{\rm eff}^{\mu\nu}\xi_\mu\xi_\nu$ is everywhere understood as taken with respect to the de Sitter metric $\bar{g}_{\mu\nu}^{\rm eff}$, and $\bar\nabla_\nu$ is the covariant derivative associated to this metric. As before, we define the covariant derivative of the Goldstone field $\xi^\mu$ by
\be
\omega_K^\mu = \omega_{\hat P}^\nu\mathcal D_\nu\xi^\mu~,
\ee
which implies
\be
\mathcal D_\nu\xi_\mu = e^{\pi}\left[\bar\nabla_\nu\xi_\mu-\left(e^\pi\xi^2+\frac{H^2}{2}\sinh\pi\right)\bar g_{\mu\nu}\right]~.
\label{xicovariantderivative}
\ee
The covariant derivative can be written explicitly in terms of $\pi$ using~(\ref{dSHiggs}) as
\be
\mathcal D_\nu\xi_\mu = \frac{1}{2}\partial_\nu\pi\partial_\mu\pi-\frac{1}{2}\bar\nabla_\nu\bar\nabla_\mu\pi-\frac{1}{4}\bar g^{\alpha\beta}\partial_\alpha\pi\partial_\beta\pi\bar g_{\mu\nu}-\frac{H^2}{4}e^{2\pi}\bar g_{\mu\nu}+\frac{H^2}{4}\bar g_{\mu\nu}~.
\label{dsdmuxinu}
\ee
The other key ingredient for writing down invariant actions is the metric. Noting that the appropriate vielbein is $e_\mu^m = e^\pi \bar e_\mu^m$, we see that the appropriate metric with which to contract indices is
\be
g_{\mu\nu}^{\rm eff} = e^{2\pi}\bar g_{\mu\nu}^{\rm eff}\,.
\ee
Finally, the invariant volume element is given by
\be
\frac{1}{4!}\epsilon_{\mu\nu\rho\sigma}\omega_{\hat P}^\mu\wedge\omega_{\hat P}^\nu\wedge\omega_{\hat P}^\rho\wedge\omega_{\hat P}^\sigma =  \rd^4y\sqrt{-\bar g_{\rm eff}}~e^{4\pi}= \rd^4y\sqrt{-g_{\rm eff}}~.
\label{dsinvvolume}
\ee
Although expressed in terms of $y^\mu$ coordinates, the answer is manifestly diffeomorphism invariant and hence holds in any coordinate system. 

The Goldstone action is then formed by building scalars from these ingredients. (As before we are allowed to use the matter covariant derivative, $\nabla_\mu$ -- the covariant derivative associated to $g_{\mu\nu}^{\rm eff}$ -- but for the lowest order actions we will not need it.) The simplest action is just the conformally invariant volume (\ref{dsinvvolume}), analogous to~(\ref{S0Poincare}),
\be
S_0 = M_{\rm v}^4\int\rd^4y\sqrt{-\bar g_{\rm eff}}e^{4\pi} \,.
\ee
Meanwhile, the kinetic term for the Goldstone field arises from\footnote{Incidentally,
$S_1$ can be realized as a wedge product as follows
\be
\nonumber
S_1 = \frac{M_0^2}{3!} \int\epsilon_{\mu\nu\rho\sigma}\omega_K^\mu\wedge\omega_{\hat P}^\nu\wedge\omega_{\hat P}^\rho\wedge\omega_{\hat P}^\sigma~.
\label{dsl1}
\ee
}
\begin{align}
\nonumber
S_1 = -M_0^2\int\rd^4y\sqrt{-g_{\rm eff}}\mathcal D_\mu\xi^\mu &=M_0^2\int\rd^4y\sqrt{-\bar g_{\rm eff}}\left[\frac{1}{2}e^{2\pi}(\partial\pi)^2+\frac{1}{2}e^{2\pi}\bar \square\pi-H^2e^{2\pi}+H^2e^{4\pi}\right]\\
&= M_0^2\int\rd^4y\sqrt{-\bar g_{\rm eff}}\left[-\frac{1}{2}e^{2\pi}(\partial\pi)^2 - H^2e^{2\pi}+H^2e^{4\pi}\right]~,
\label{s1ds}
\end{align}
where all contractions are performed with the de Sitter metric $\bar{g}_{\mu\nu}^{\rm eff}$ and in the last line we have integrated by parts. Note that this expression has a tadpole contribution which may be canceled by adding a suitable multiple of the invariant measure, thereby setting the relative coefficient between $S_1$ and $S_0$. As a check, this result agrees with~(\ref{S1Poincare}) in the limit $H\rightarrow 0$. 

At the four-derivative level, we have\footnote{As before, this term may also be constructed directly as a wedge product of Maurer--Cartan coefficients:
\be
\nonumber
S_2 = -\int \left(\frac{1}{2}\epsilon_{\mu\nu\rho\sigma}\omega_K^\mu\wedge\omega_K^\nu\wedge\omega_{\hat P}^\rho\wedge\omega_{\hat P}^\sigma + \bar g_{\mu\nu}^{\rm eff}\omega_K^\mu\wedge\star_4\omega_K^\nu\right)~.
\ee}
\bea
\nonumber
S_2 &=& \int \rd^4y\sqrt{-g_{\rm eff}}\left(\mathcal D_\mu\xi^\mu\right)^2\\
&=& \frac{1}{4}\int \rd^4y\sqrt{- \bar g_{\rm eff}}\bigg[(\bar\square\pi)^2+2\bar\square\pi(\partial\pi)^2+(\partial\pi)^4-4H^2(\partial\pi)^2\bigg]-\frac{8H^2}{M_0^2} S_1-\frac{4H^4}{M_{\rm v}^4}S_0~,
\label{dmumu}
\eea
where we have dropped a total derivative and a constant ($\pi$-independent) term. The last two terms can of course be absorbed
into the coefficients of the lower-order action $S_0$ and $S_1$. 

There is of course another four-derivative term, obtained from $\left(\mathcal D_\mu\xi_\nu\right)^2$, but the corresponding action turns out to be a linear combination of $S_2$, $S_1$ and $S_0$:
\be
S_{2}' = \int{\rm d}^4y\sqrt{-g_{\rm eff}}\left(\mathcal D_\mu\xi_\nu\right)^2 =  -\int \bar g_{\mu\nu}^{\rm eff}\omega_K^\mu\wedge\star_4\omega_K^\nu  = S_2 + \frac{6H^2}{M_0^2}S_1 + \frac{3H^4}{M_{\rm v}^4}S_0~.
\label{dmunusq}
\ee
However, this degeneracy is an accident of $d = 4$ dimensions. One can form a linearly independent combination of these two terms in $d$-dimensions and then take the limit $d\to 4$ in order to recover another invariant combination~\cite{Nicolis:2008in}. The result of this procedure, detailed in Sec.~\ref{curvinvariants}, is the orthogonal combination
\be
S_{\rm wz} = \int \rd^4y\sqrt{-\bar g_{\rm eff}}\Big[(\partial\pi)^4+2\bar\square\pi(\partial\pi)^2-6H^2(\partial\pi)^2-12H^4\pi\Big]~.
\label{SWZ1}
\ee
As indicated by the subscript, this is a Wess--Zumino (WZ) term, in the same sense as $\mathcal L_3$ of the conformal galileons \cite{Nicolis:2008in,Goon:2012dy}.\footnote{The WZ term can be constructed in five dimensions as 
$ S_{\rm wz}= \int_M\epsilon_{\mu\nu\rho\sigma}\omega_D\wedge\omega_K^\mu\wedge\omega_K^\nu\wedge\omega_{\hat P}^\rho\wedge\omega_{\hat P}^\sigma$, and then pulled back to the physical four-dimensional space-time using Stokes' theorem~\cite{Goon:2012dy}.} Notice that this term contains a kinetic term for the field $\pi$, but no corresponding mass term, which would seem to contradict the statement that the Goldstone field has a fixed mass. However, this term also contains a tadpole, which must be cancelled off by adding a suitable amount of the $e^{4\pi}$ potential. This relative tuning then leads to a mass term with $m^2 = -4H^2$, as expected.\footnote{Thanks to James Bonifacio for pointing out an error in the previous version of~\eqref{SWZ1}.}

The construction of the effective action can be extended in this way to arbitrary derivative order.  To summarize, the most general Goldstone lagrangian consistent with the symmetry 
breaking pattern~(\ref{symbreakbis}), up to fourth order in derivatives, is\footnote{Note that the $M_1$ and $M_2$ higher-derivative terms include $H^2(\partial\pi)^2$ corrections to the kinetic term,
which were not included in the two-derivative analysis of~\cite{Hinterbichler:2011qk}. However, in order for the effective field theory paradigm to be useful, we are assuming that there is a hierarchy of scales such that the higher-order terms are sub-dominant, {\it i.e.}, $M_{1,2} \ll H^2$. The benefit of this approach is that it allows us to systematically include the effects of such corrections, but for the time being we ignore them.}
\begin{align}
\label{Goldstonelag}
\nonumber
S_\pi = \int {\rm d}^4x \sqrt{-\bar g_{\rm eff}}&\bigg[M_0^2\left(-\frac{1}{2}e^{2\pi}(\partial\pi)^2- H^2e^{2\pi}+ \frac{H^2}{2}e^{4\pi}\right)\\
&~~~~~+M_1\bigg((\bar\square\pi)^2+2\bar\square\pi(\partial\pi)^2+(\partial\pi)^4-4H^2(\partial\pi)^2\bigg)\\
\nonumber
&~~~~~+M_2\bigg((\partial\pi)^4+2\bar\square\pi(\partial\pi)^2-6H^2(\partial\pi)^2-12H^2\pi+3H^2e^{4\pi}\bigg)+\ldots\bigg]\,,
\end{align}
where the relative coefficient between the $e^{2\pi}$ and $e^{4\pi}$ terms has been fixed to cancel the $\pi$ tadpole.

\subsection{Transformation of $\pi$}
Up to this point, we have not specified how $\pi$ transforms under the non-linearly realized conformal symmetries, though it is implicit in the construction.  A straightforward way to determine this transformation rule explicitly is to act on the left of (\ref{dscosetsection}) by a group element, $\bar g \in $ G, and determine how $\pi$ transforms.  Note that this will be tied to a particular coordinitization of de Sitter space.

There is, in fact, a simpler method to derive the transformation rule for $\pi$ in a coordinate-independent way. This method is closely tied to a technique we will use in Sec.~\ref{curvinvariants} as an alternative
to the coset construction. Consider the metric $g_{\mu\nu}^{\rm eff} = e^{2\pi}\bar g_{\mu\nu}^{\rm eff} $, where $\bar g_{\mu\nu}^{\rm eff} $ is the Sitter metric in an arbitrary coordinate system. Clearly $g_{\mu\nu}^{\rm eff}$
non-linearly realizes the conformal group through the dilaton field $\pi$. We can extract the transformation properties for the scalar mode $\pi$ from the general transformation properties of the metric under an
infinitesimal diffeomorphism
\bea
\nonumber
\delta g_{\mu\nu}^{\rm eff} &=& -g_{\rho\nu}^{\rm eff}\nabla_{\mu}\xi^{\rho}-g_{\mu\rho}^{\rm eff}\nabla_{\nu}\xi^{\rho}\\
&=& 2\delta\pi  e^{2\pi}\bar g_{\mu\nu}^{\rm eff} \,,
\eea
where in the last step we have restricted our attention to conformal transformations. Taking the trace gives the desired transformation rule for $\pi$:
\be
\delta\pi = -\xi^\mu\partial_\mu\pi - \frac{1}{4}\bar\nabla_\mu\xi^\mu~.
\label{delpi}
\ee
This displays all the desired properties. Under $\mathfrak{so}(4,1)$ transformations, corresponding to isometries of $\bar g_{\mu\nu}^{\rm eff}$, $\pi$ transforms linearly.
Under the other conformal transformations, $\pi$ transforms non-linearly. 

Given a particular coordinatization of de Sitter, the infinitesimal transformations for the fields can be worked out explicitly. For example, in the flat slicing $\bar g_{\mu\nu}^{\rm eff} = H^{-2}t^{-2}\eta_{\mu\nu}$,~(\ref{delpi}) gives
\begin{align}
\nonumber
\delta_{P_\mu}\pi &= -\partial_\mu\pi +\delta_{\mu}^0\frac{1}{t},\\
\nonumber
\delta_{J_{\mu\nu}}\pi &=  (x_\mu\partial_\nu - x_\nu\partial_\mu)\pi + \left(\delta^0_\mu\frac{x_\nu}{t}-\delta^0_\nu\frac{x_\mu}{t}\right),\\
\nonumber
\delta_D\pi &= -x^\mu\partial_\mu\pi,\\
\delta_{K_\mu}\pi &= -(2x_\mu x^\nu\partial_\nu-x^2\partial_\mu)\pi -\delta_\mu^0\frac{x^2}{t}~.
\end{align}
Consistent with the discussion of Sec.~\ref{intro}, the symmetries associated to $P_0, K_0$ and $J_{0i}$ are non-linearly realized, while the others are linearly realized.
Furthermore, $\pi$ transforms as a weight 0 field under dilations.

\subsection{Matter Fields}
In the pseudo-conformal scenario, the progenitor of density perturbations is not the Goldstone field $\pi$ associated with the time-evolving field, but rather a weight-0 spectator field, $\chi$.
As a result, we need to couple matter fields to the Goldstone in a way that non-linearly realizes the conformal group. Of course, the coset machinery is also capable of this task. 

Recall that the covariant derivative of a matter field is given by
\be
\omega_{\hat P}^\mu \bar{\mathcal D}_\mu\chi= {\rm d}\chi+\omega_V^iD(V_i)\chi+{1\over 2}\omega_J^{\mu\nu}D(J_{\mu\nu})\chi~.
\ee
For this symmetry-breaking pattern, there are no elements, $\omega_V$, of the Maurer--Cartan form that play the role of a gauge connection, so we only need to concern ourselves with the spin connection piece $\omega_J$. Note under the Weyl transformation
\be
\tilde e_\mu^m = e^\pi \bar e_\mu^m~,
\ee
the spin connection transforms as
\be
\widetilde\omega^{mn}_\mu =\omega^{mn}_\mu + e^n_\mu\partial^m\pi - e^m_\mu\partial^n\pi~.
\ee
Thus the spin connection (\ref{dsmcform1}) is in fact the spin connection associated to the metric $g_{\mu\nu}^{\rm eff} = e^{2\pi}\bar g_{\mu\nu}^{\rm eff} $, where $\bar g_{\mu\nu}^{\rm eff} $ is a metric on de Sitter space. In other words, the covariant derivative for $\chi$ is just the geometric covariant derivative associated to this metric
\be
\bar{\mathcal D}_\mu\chi = \nabla_\mu\chi~.
\ee

An action for $\chi$ can be obtained by contracting indices with the conformal metric, $g^{\rm eff}_{\mu\nu}$, which will introduce a natural coupling between $\chi$ and $\pi$.
In particular, because $\chi$ is a weight 0 field, there is the additional freedom to promote any of the mass scales in the Goldstone lagrangian~(\ref{Goldstonelag}) to a function of $\chi$,
being careful about integration by parts. (An important exception is the Wess--Zumino term~(\ref{SWZ1}). This term shifts by a total derivative under conformal transformations, hence 
its coefficient must remain independent of $\chi$.)

With these caveats in mind, we are free to write down any Lorentz-invariant action using $\chi$, the effective metric $g_{\mu\nu}^{\rm eff}$ and its covariant derivative $\nabla_\mu$. At the end of the day, the result can be expressed in terms of the effective de Sitter metric $\bar g_{\mu\nu}^{\rm eff}$. 
Here are some fiducial terms in the effective lagrangian for $\chi$ (written in terms of the effective de Sitter metric $\bar g_{\mu\nu}^{\rm eff}$):
\begin{align}
\label{weight0action}
\nonumber
S_\chi = \int\rd^4x\sqrt{-\bar g_{\rm eff}}&\left[-\frac{\bar M^2_\chi}{2}e^{2\pi}(\partial\chi)^2+e^{4\pi}V(\chi)+a_1(\partial\chi)^4+a_2(\bar\square\chi)^2+\ldots\right.\\
&+\bar M^2_0(\chi)\left(\frac{1}{2}e^{2\pi}(\partial\pi)^2+\frac{1}{2}e^{2\pi}\bar\square\pi-H^2e^{2\pi}+\frac{H^2}{2}e^{4\pi}\right)\\
\nonumber
&+\left.\bar M_1(\chi)\bigg((\bar\square\pi)^2+2\bar\square\pi(\partial\pi)^2+(\partial\pi)^4-4H^2(\partial\pi)^2\bigg)+\ldots\right]~,
\end{align}
where the ellipses in the first line indicates higher order terms in $\chi$ with no derivatives on $\pi$. We have additionally assumed that $\chi$ is canonically normalized, up to an overall constant mass scale $M_\chi^2$. Furthermore, we assume that $V(0)=\rd V/\rd\chi\rvert_0=\rd M_0^2/\rd\chi\rvert_0 = 0$ so that there are no tadpole terms for either $\pi$ or $\chi$.
\section{Method of Curvature Invariants}
\label{curvinvariants}
The coset construction machinery of the previous sections, while extremely powerful, is technically involved, hence it is pedagogically helpful to present an alternative way of
deriving our effective lagrangians. The technique is an extension of the method used in \cite{Nicolis:2008in} to obtain the conformal galileon combinations, which we foreshadowed
in deriving the transformation rule for $\pi$ in the last section. 

The basic idea is the following. To linearly realize the de Sitter group, ${\rm SO}(4,1)$, our theory should be cast in terms of a (fictitious) de Sitter metric,
$\bar g_{\mu\nu}^{\rm eff}$, and its covariant derivative. In addition, we also want to non-linearly realize the conformal group ${\rm SO}(4,2)$. This is achieved by introducing
the conformal mode:
\be
g_{\mu\nu}^{\rm eff} =e^{2\pi}\bar g_{\mu\nu}^{\rm eff}~.
\label{g}
\ee
This metric is clearly conformally invariant, with $\pi$ transforming in some non-linear fashion under a general conformal transformation. 
To simplify the notation, we will omit the subscript ``eff", with the implicit understanding that all metrics in the effective theory are
fictitious. 

By using the geometric covariant derivative associated to this conformal metric, we can write down invariant actions for matter fields that non-linearly realize the conformal group. In order to get the action for the Goldstone we want to consider curvature invariants, which pick out the dynamics of the conformal mode $\pi$. To see that this method is completely equivalent to the coset construction, first note that
because the metric~(\ref{g}) is obviously conformal to de Sitter -- and thus conformally flat -- all of the curvature information is contained in the Ricci tensor
\be
R_{\mu\nu} =  3H^2\bar g_{\mu\nu} - 2\bar\nabla_\mu\bar\nabla_\nu\pi-\bar g_{\mu\nu}\bar\square\pi+2\partial_\mu\pi\partial_\nu\pi-2\bar g_{\mu\nu}(\partial\pi)^2~,
\label{dsricci}
\ee
where all derivatives and contractions are with respect to the background de Sitter metric $\bar g_{\mu\nu}^{\rm eff}$. It is possible to write $R_{\mu\nu}$ in terms of (\ref{dsdmuxinu}) as
\be
R_{\mu\nu} = 4\mathcal D_\mu\xi_\nu+2\mathcal D_\alpha\xi^\alpha g_{\mu\nu}+3H^2 g_{\mu\nu}~.
\ee
Tracing over this, it is possible to express the Ricci scalar as
\be
R = 12\mathcal D_\mu\xi^\mu+12H^2~.
\ee
Additionally, we know that the covariant derivative associated to $g_{\mu\nu}$ is a building block in both cases. Therefore we see that the building blocks for the curvature invariant story $\left\{g_{\mu\nu},~R_{\mu\nu},~\nabla_\mu\right\}$, are equivalent to the ingredients of the coset construction $\left\{g_{\mu\nu},~\mathcal D_\mu\xi_\nu,~\nabla_\mu\right\}$. The curvature invariant prescription therefore provides an equivalent, and less technically demanding, route to build invariant actions.  
\subsection{Goldstone Actions}
\label{confpiaction}
The algorithm for constructing invariant actions of the Goldstone is extremely simple: we may construct any diffeomorphism scalar from the conformal metric, Ricci tensor and covariant derivatives of $g_{\mu\nu}$. The simplest invariant is of course just the invariant measure
\be
S_0 \sim \int\rd^4x\sqrt{-g} =\int\rd^4x\sqrt{-\bar g}e^{4\pi}~.
\ee
The kinetic term for the field $\pi$ comes from the Ricci scalar
\be
S_1 \sim \int\rd^4x \sqrt{-g}\left[ -R + 6H^2\right] \sim
\int\rd^4x \sqrt{-\bar g}\left[\frac{1}{2}e^{2\pi}(\partial\pi)^2+\frac{1}{2}e^{2\pi}\bar\square\pi -H^2e^{2\pi}+ \frac{H^2}{2}e^{4\pi}\right]~,
\label{fakeaction}
\ee
where we have explicitly included the cosmological term in order to cancel the tadpole, much as in the previous section. This agrees with~(\ref{s1ds}), up to the $S_0$ term.

At the four-derivative level, we can consider $R^2$ and $R_{\mu\nu}^2$. As in the flat space case, these terms both give the same action for $\pi$ (after integration by parts):
\be
S_2 \sim \int\rd^4x\sqrt{-g} \left(R^2,\,R_{\mu\nu}^2\right) \sim \int \rd^4 x\sqrt{-\bar g}\Big[(\bar\square\pi)^2+2\bar\square\pi(\partial\pi)^2+(\partial\pi)^4-4H^2(\partial\pi)^2\Big]~,
\label{scalarsq}
\ee
where we have discarded a total derivative and a constant. Note that this agrees with (\ref{dmumu}) up to terms which are multiples of $S_1$ and $S_0$.
This degeneracy of the $R^2$ and $R_{\mu\nu}^2$ terms is an ``accident" of $d=4$, as was noted for the conformal galileon case in~\cite{Nicolis:2008in}.
We observed a similar phenomenon from the coset construction; (\ref{dmumu}) and (\ref{dmunusq}) had the same form at highest order in derivatives. 

However, by constructing these terms in $d$-dimensions and then take a suitable limit, we can obtain another independent linear combination~\cite{Nicolis:2008in}. Consider the linear combination
\begin{align}
\nonumber
\sqrt{- g}\left(\frac{R_{\mu\nu}^2}{(d-1)}-\frac{R^2}{(d-1)^2}\right)=&~e^{(d-4)\pi}\bigg[-dH^4+2(d-1)H^2\bar\square\pi+(d-4)^2H^2(\partial\pi)^2+(d-4)(\bar\square\pi)^2\\
&\hspace{-.1cm}+\frac{(d-4)(d-2)(3d-4)}{2(d-1)}\bar\square\pi(\partial\pi)^2+\frac{(d-4)(d-2)^3}{2(d-1)}(\partial\pi)^4\bigg]~.
\end{align}
The source of the degeneracy is made manifest by the fact that this combination is proportional to $d-4$. To get a non-vanishing result as $d\to 4$, we should therefore divide by this factor before taking the limit:
\be
\lim_{d\rightarrow 4}\frac{\sqrt{-g}}{(d-4)}\left[\frac{R_{\mu\nu}^2}{(d-1)}-\frac{R^2}{(d-1)^2}\right]= \sqrt{-\bar g}\left[(\bar\square\pi)^2+\frac{8}{3}\bar\square\pi(\partial\pi)^2+\frac{4}{3}(\partial\pi)^4-6H^2(\partial\pi)^2-4H^4\pi\right]~.
\ee
By taking a suitable linear combination of this term and~(\ref{scalarsq}), we can form the Wess--Zumino combination, already given in~(\ref{SWZ1}),
\be
S_{\rm wz} = \int \rd^4x\sqrt{-\bar g}\Big[(\partial\pi)^4+2\bar\square\pi(\partial\pi)^2-6H^2(\partial\pi)^2-12H^4\pi\Big]~.
\label{SWZ}
\ee
This term is a Wess--Zumino term in the sense that it cannot be constructed via the coset construction in $d=4$.  But it is not a Wess--Zumino term in other dimensions, which is why the limiting procedure above works.   Note that~(\ref{SWZ}) reduces to
the cubic conformal galileon $\mathcal L_3$ in the limit $H \to 0$. 
\subsection{Adding Matter Fields}
Introducing matter fields is straightforward in this formalism. It is clear that any diffeomorphism invariant action constructed from matter fields and the
conformal metric~(\ref{g}) will linearly realize the de Sitter group while non-linearly realizing the conformal group. Consider for concreteness a weight zero field $\chi$.
Any scalar built from $\chi$, the conformal metric and its covariant derivative will necessarily enjoy the desired symmetries, for instance
\be
S_\chi = \int \rd^4x\sqrt{-g}\left[-\frac{M_\chi^2}{2}(\partial\chi)^2 - \frac{m_\chi^2}{2}\chi^2+\lambda_\chi\chi^3+\ldots\right]~,
\ee
where $g_{\mu\nu} =e^{2\pi}\bar g_{\mu\nu}$. More generally, because $\chi$ has weight 0, we are allowed to construct curvature scalars from the metric as in Sec.~\ref{confpiaction} and promote their coefficients to a polynomial function of $\chi$
\begin{align}
\nonumber
S_\chi = \int\rd^4x\sqrt{-g}&\left[-\frac{M_\chi^2}{2}(\partial\chi)^2+V(\chi)+a_1(\partial\chi)^4+a_2(\square\chi)^2+\ldots\right.\\
+&\bar M_0^2(\chi) R + \bar M_1(\chi) R^2 + \bar M_3(\chi)R^3 +\bar M_4(\chi)R\square R+\bar M_5(\chi)R_{\mu\nu}^3+\ldots\bigg]
\end{align}
As mentioned in Sec.~\ref{conftods}, however, the Wess--Zumino term~(\ref{SWZ}) is an important exception. Because the corresponding lagrangian density shifts by a total derivative, it cannot be multiplied by an arbitrary function of $\chi$ without explicitly breaking conformal invariance.
\section{Analysis of the Low Energy Effective Action}
\label{lowenergyeffectiveaction}
Although a more thorough analysis of the effective lagrangian is underway, here we focus on analyzing its elementary properties. For concreteness, we work up 
to cubic order in the field $\chi$, and to second order in derivatives. With these simplifications, the combined action $S_\pi+S_\chi$ given by (\ref{Goldstonelag}) and (\ref{weight0action}) reduces to
\begin{align}
\nonumber
S = \int {\rm d}^4x \sqrt{-\bar g_{\rm eff}}\bigg[M_0^2\bigg(-\frac{1}{2}e^{2\pi}&(\partial\pi)^2- H^2e^{2\pi}+ \frac{H^2}{2}e^{4\pi}\bigg)-\frac{ M^2_\chi}{2}e^{2\pi}(\partial\chi)^2+\frac{m_\chi^2}{2}e^{4\pi}\chi^2+\lambda_\chi e^{4\pi}\chi^3\\
&+ \bar M^2_0\left(\frac{1}{2}e^{2\pi}(\partial\pi)^2+\frac{1}{2}e^{2\pi}\bar\square\pi-H^2e^{2\pi}+\frac{H^2}{2}e^{4\pi}\right)(\chi^2+\alpha\chi^3)\bigg] \,,
\label{Spichi}
\end{align}
where all contractions are with respect to the de Sitter metric $\bar g^{\rm eff}_{\mu\nu}$. In obtaining this result, we have Taylor-expanded the function $\bar M_0(\chi)$ in~(\ref{weight0action}) to third order in fields, with $\alpha$ denoting a dimensionless constant. By assumption, there is no linear term in $\chi$, as discussed below~(\ref{weight0action}).

\subsection{Two-Point Function for the Goldstone}
First we consider the two-point function for the Goldstone mode $\pi$. The quadratic action for $\pi$ that derives from~(\ref{Spichi}) is
\be
S_{\pi} = M_0^2\int \rd^4x\sqrt{-\bar g_{\rm eff}}\left[-\frac{1}{2}(\partial\pi)^2+2H^2\pi^2\right]~.
\ee
To proceed, we must choose a coordinatization of de Sitter. A convenient choice is the flat slicing
\be
\rd s^2 = \frac{1}{H^2t^2}\left(-\rd t^2+\rd \vec x^2\right)~.
\label{flatdesitter}
\ee
Here we have written the conformal time coordinate as $t$ because it is really the physical Minkowski space-time coordinate, it merely acts as a conformal time coordinate on the effective de Sitter space that spectator fields feel. In terms of this metric, the action takes the form
\be
S_\pi = M_0^2\int \rd^4 x\left[\frac{1}{2H^2t^2}\dot\pi^2-\frac{1}{2H^2t^2}(\vec\nabla\pi)^2+\frac{2}{H^2t^4}\pi^2\right]~.
\ee
The equation of motion for the $\pi$ field is given in Fourier space by
\be
\ddot\pi_k +k^2\pi_k -\frac{2}{t}\dot\pi_k-\frac{4}{t^2}\pi_k = 0 
\ee
After performing a field redefinition to the canonically-normalized variable, $v = \frac{M_0}{(-Ht)}\pi$, the mode function equation becomes
\be
\ddot v_k +\left(k^2-\frac{6}{t^2}\right)v_k = 0~.
\ee
Assuming adiabatic vacuum initial conditions, it is well-known that this equation admits a solution in terms of a Hankel function of the first kind
\be
v_k(t) = \sqrt{\frac{\pi(-t)}{4}}H_{5/2}^{(1)}(-kt)~.
\ee
Inverting our field redefinition to get an expression for $\pi$ we find
\be
\pi_k(t) = -i\frac{H(-t)^{3/2}}{M_0}\sqrt\frac{\pi}{4}H^{(1)}_{5/2}(-kt) = \frac{-3H}{\sqrt{2 k^5}(-t)M_0}\left(1+ikt-\frac{k^2t^2}{3}\right)e^{-ikt}
\ee
Using the asymptotic expansion for the Hankel function, $H^{(1)}_{5/2}(x) \sim -3i\sqrt{2/\pi}x^{-5/2}$ for $x\ll 1$,
the long-wavelength ($\lvert kt\rvert \ll 1$) power spectrum for $\pi$ is
\be
\mathcal P_\pi = \frac{1}{2\pi^2}k^3\lvert\pi_k\rvert^2 \sim \frac{9H^2}{(2\pi)^2M_0^2}\frac{1}{(-kt)^2}~.
\ee
Note that this spectrum peaks at long wavelengths and is thus strongly red-tilted.
\subsection{Two-Point Function for Massless Spectator Fields} 
Now let us compute the power spectrum for the weight-0 spectator field $\chi$. Recall that this is the field that we envision will lead to a scale-invariant spectrum of curvature perturbations once these entropic perturbations have been converted to the adiabatic direction. A detailed analysis of the conversion of perturbations is beyond the scope of this paper, but is the subject of current work. 

At quadratic order in $\chi$, the action~(\ref{Spichi}) gives
\be
S_\chi = \int \rd^4x\sqrt{-\bar g_{\rm eff}}\left[-\frac{M_\chi^2}{2}(\partial\chi)^2-\frac{m_\chi^2+\bar M^2_0H^2}{2}\chi^2\right]~,
\ee
which just describes a massive scalar field on de Sitter space. It is well-known that the field will acquire a scale-invariant spectrum of fluctuations provided that its mass is sufficiently small:
$m_\chi^2/(M_\chi^2H^2)$ and $\bar M^2_0/M_\chi^2\ll 1$. Indeed, ignoring the mass term, the solution for the canonically normalized variable $\hat\chi = \frac{M_\chi}{(-Ht)}\chi$ is
\be
\hat\chi_k = \frac{1}{\sqrt{2k}}\left(1-\frac{i}{kt}\right)e^{-ikt}~,
\ee
where the usual adiabatic vacuum has been assumed. This implies that the long-wavelength power spectrum for $\chi_k$ is
scale invariant
\be
\mathcal P_\chi = \frac{1}{2\pi^2}k^3\lvert\chi_k\rvert^2 \sim \frac{H^2}{(2\pi)^2M_\chi^2}~.
\ee

\subsection{Three-Point Function Involving Massless Spectator Fields}
The pseudo-conformal scenario is extremely constrained by symmetry --  we now investigate how the symmetries at play affect some of the three point functions of the theory. For simplicity, we work in the exactly scale-invariant limit for 
$\chi$, corresponding to choice $m_\chi = \bar M_0 = 0$. Up to cubic order in the fields, the action~(\ref{Spichi}) then gives
\bea
\nonumber
S_3 &=&  \int \rd^4x\sqrt{-\bar g_{\rm eff}}\left[-\frac{M_0^2}{2}(1+2\pi) (\partial\pi)^2 + 2H^2M_0^2\pi^2 + 4M_0^2H^2\pi^3\right]\\
&+& \int \rd^4x\sqrt{-\bar g_{\rm eff}}\left[-\frac{M_\chi^2}{2}(\partial\chi)^2-M_\chi^2\pi(\partial\chi)^2+\lambda_\chi\chi^3\right]~.
\label{3pointaction}
\eea
We will use this action to compute the $\langle \chi\chi\chi\rangle$ and $\langle \pi\chi\chi\rangle$ correlation functions at tree-level. The $\langle\chi^3\rangle$ correlation function is interesting because it will contribute to the non-gaussian signature of the field $\zeta$ after conversion; while the $\langle\pi\chi^2\rangle$ correlator is important because mixing of $\pi$ with the spectator field $\chi$ is what non-linearly realizes conformal symmetry. We therefore expect this correlation function to have non-trivial properties under conformal variation.

\noindent {\bf $\chi\chi\chi$ Three-Point Function}

The $\langle \chi^3\rangle$ correlation function follows at tree-level from the $\lambda_\chi\chi^3$ vertex. Note that this vertex is the most general
cubic interaction term we need to consider, up to field redefinitions. Indeed, even though we can imagine an infinite number of higher-derivative
cubic interactions, such as $\chi\bar{\nabla}^\mu\bar{\nabla}^\nu\chi\bar{\nabla}_\mu\bar{\nabla}_\nu\chi$, these can all be brought to the form $\chi^3$ through integration by parts and
suitable field redefinitions~\cite{Creminelli:2011mw}.

We can now compute the $\langle\chi^3\rangle$ correlator using the standard in-in formalism \cite{Maldacena:2002vr, Weinberg:2005vy} after choosing the flat slicing of de Sitter (\ref{flatdesitter}) for our background metric. At tree level the equal time correlator is given by
\be
\langle\chi_{k_1}\chi_{k_2}\chi_{k_3}\rangle = -i\int_{-\infty}^{t}\rd t'\langle 0\rvert\left[\chi_{k_1}(t)\chi_{k_2}(t)\chi_{k_3}(t), ~H_{\rm int}(t')\right]\lvert 0\rangle~,
\ee
where
\be
H_{\rm int}(t') = -\int \rd^3x \mathcal L_{\rm int} = -\lambda_\chi\int \rd^3x\frac{1}{H^4{t'}^4}\chi^3~.
\ee
Performing the computation reproduces the well-known result~\cite{Creminelli:2011mw, Zaldarriaga:2003my}
\be
\langle\chi_{k_1}\chi_{k_2}\chi_{k_3}\rangle = (2\pi)^3\delta^{(3)}(\vec k_1+\vec k_2+\vec k_3)\frac{\lambda_\chi H^2}{2M_\chi^6}\frac{1}{\prod_i k_i^3}\left[k_1k_2k_3 -\sum_{i\neq j}k_i^2k_j-\sum_i k_i^3\left(1-\gamma-\log k_t t_*\right)\right]~,
\label{3ptchi}
\ee
where $k_t = k_1 + k_2 + k_3$, $\gamma$ is the Euler gamma, and $t_*$ is a late time cutoff introduced to regulate the infrared divergence of the integral. Not surpringly, this result is identical to the three-point function for a massless spectator field in inflation~\cite{Creminelli:2011mw}. This is due to the fact that the correlator is invariant under $\mathfrak{so}(4,1)$ symmetries in both cases. Following~\cite{Creminelli:2011mw}, we can check this directly. At late times, the isometries of de Sitter act as the conformal group on $\mathbb R^3$. In momentum space, the special conformal generators act on correlators as \cite{Creminelli:2011mw, Maldacena:2011nz}
\be\label{confsymgen}
\delta = \sum_a 2(3-\Delta_a)\vec b\cdot\vec\partial_{k_a}-\vec b\cdot \vec k_a\vec\partial^2_{k_a}+2\vec k_a\cdot\vec\partial_{k_a}(\vec b\cdot\vec\partial_{k_a})~,
\ee
where $a$ indexes the fields in the correlator and the $3d$ conformal weight, $\Delta$, of a mass $m$ field is defined as
\be
\Delta = \frac{3}{2}-\sqrt{\frac{9}{4}-\frac{m^2}{H^2}}.
\ee
Then, one can check explicitly that (\ref{3ptchi}) is annihilated by this operator with $\Delta_a = 0$ for $a = 1, 2, 3$. This is a nice consistency check since the action we have constructed
linearly realizes the de Sitter group. Note that there does not appear to be a further restriction on the amplitude coming from the requirement of non-linearly realizing the conformal symmetry. Intuitively this makes sense, as we expect conformal invariance to be a result of mixing with the dilaton-like $\pi$ field.

\noindent {\bf $\pi\chi\chi$ Three-Point Function}

As another example, we consider computing the equal time correlation function for the $\pi\chi^2$ interaction in (\ref{3pointaction}). In this case, the interaction Hamiltonian is given by
\be
H_{\rm int}(t') = -\int \rd^3x \mathcal L_{\rm int} = \int \rd^3x\frac{1}{H^2{t'}^2}\left[-\pi\dot\chi^2+\pi(\vec\nabla\chi)^2\right]~.
\ee
Computing the equal-time correlator by standard techniques, we obtain the result
\be
\langle\pi_{k_1}\chi_{k_2}\chi_{k_3}\rangle = (2\pi)^3\delta^{(3)}(\vec k_1+\vec k_2+\vec k_3)\frac{3\pi H^4}{16M_0^2M_\chi^2}\frac{1}{k_1^5k_2^3k_3^3 t_*}\bigg[k_1^4+2k_1^2(k_2^2+k_3^2)-3(k_2^2-k_3^2)^2\bigg]~.
\ee
As before, $t_*$ is a late-time cutoff introduced to regulate the infrared divergence of the correlator. Note that for the field $\pi$, $m_\pi^2 = -4H^2$ so it has de Sitter weight ($3d$ conformal weight) $\Delta = -1$. Therefore, the time dependence of this correlator is to be expected, as fields scale as $\phi \sim t^\Delta$ at late times in de Sitter. It can again be checked that this correlator is invariant under the $\mathfrak{so}(4,1)$ symmetry generator (\ref{confsymgen}).
\subsection{Constraints From Symmetry}
In the examples considered so far, the correlators are constrained to be invariant under the linearly-realized $\mathfrak{so}(4,1)$ symmetries. This is identical to the picture in inflation, where spectator fields obey the same symmetry algebra. Indeed, the constraint that the de Sitter group acts as the conformal group on spatial slices at late times constrains the real-space three point function for any spectator fields to be of the form \cite{DiFrancesco:1997nk}
\be
\lim_{t\to0}~\langle \varphi_1(\vec{x}_1,t) \varphi_2(\vec{x}_2,t)\varphi_3(\vec{x}_3,t)\rangle = \frac{C_{123}}{x_{12}^{\Delta_1 + \Delta_2-\Delta_3}x_{23}^{\Delta_2 + \Delta_3-\Delta_1}x_{13}^{\Delta_1 + \Delta_3-\Delta_2}}~.
\ee
For instance, the $\langle\pi\chi\chi\rangle$ correlator considered in the previous subsection in real space is at late times given by
\be
\langle\pi(\vec x_1, t)\chi(\vec x_2, t)\chi(\vec x_3, t)\rangle\sim \frac{\lvert \vec x_1-\vec x_2\rvert\lvert \vec x_1-\vec x_3\rvert}{\lvert \vec x_2-\vec x_3\rvert}~.
\ee
However, correlators in the pseudo-conformal scenario are additionally constrained by the non-linearly realized conformal symmetries. As an example of how these symmetries can manifest themselves, consider the simplest action for a spectator field mixing with the Goldstone $\pi$, just the kinetic term
\be
S_\chi = \int\rd^4x\sqrt{-\bar g_{\rm eff}}\left[-\frac{1}{2}e^{2\pi}(\partial\chi)^2\right] = \int\rd^4x\sqrt{-\bar g_{\rm eff}}\left[-\frac{1}{2}(\partial\chi)^2-\pi(\partial\chi)^2-2\pi^2(\partial\chi)^2+\ldots\right]
\ee
This term represents an infinite number of vertices mixing $\pi$ and $\chi$, however the precise coefficients are fixed by conformal invariance. In this theory, the ratio of the four-point function to the three-point function is therefore fixed, for example. It is this additional symmetry structure beyond mere $\mathfrak{so}(4,1)$ invariance that is due to the theory originating from a spontaneously broken CFT. We are currently exploring the precise relation between various point functions using this effective field theory formalism.

\section{Conclusions}
\label{conclusion}

The pseudo-conformal scenario is based on the idea that well before the big bang, the universe is slowly evolving and approximately described by flat, Minkowski space-time.
The matter fields (or a sector of the matter fields) at that time is approximately described by a CFT, with symmetry algebra $\mathfrak{so}(4,2)$. Some of the fields in the CFT acquire specific
time-dependent expectation values, which break part of the conformal algebra but in particular preserve the dilation symmetry. The precise symmetry breaking pattern is
$\mathfrak{so}(4,2)\rightarrow \mathfrak{so}(4,1)$. The universe is slowly contracting/expanding during this phase, corresponding to $|w| \gg 1$, and hence becomes increasingly
flat, homogeneous and isotropic. Because the scale factor evolves slowly, tensor perturbations are not appreciably excited.
A detection of primordial gravitational waves, for instance through microwave background polarization measurements, would therefore rule out the scenario.

Because the unbroken $\mathfrak{so}(4,1)$ subalgebra matches the de Sitter isometries, certain fields in the CFT (specifically weight-0 fields) acquire a scale invariant spectrum of fluctuations. 
These correspond to entropy perturbations, which must later on be converted to the adiabatic channel. This scenario was originally proposed in the context of two specific realizations, namely the U(1) quartic model~\cite{Rubakov:2009np,Craps:2007ch} and the Galilean Genesis scenario~\cite{Creminelli:2010ba}. It was subsequently realized in~\cite{Hinterbichler:2011qk} that these two incarnations actually relied on the same symmetries, and that the scenario is in fact far more general. Indeed, as usual with spontaneous symmetry breaking, much of the relevant physics is determined by the symmetry breaking pattern, irrespective of the
details of the microphysical realization. 

In this paper we applied the well-known coset construction to derive the most general effective lagrangian describing the Goldstone field $\pi$ and matter fields.
The resulting action captures the low energy dynamics of {\it any} pseudo-conformal realization, including the quartic model and Galilean Genesis. As it should, the
effective theory thus constructed linearly realizes the unbroken $\mathfrak{so}(4,1)$ symmetry and non-linearly realizes the conformal algebra $\mathfrak{so}(4,2)$.
An important subtlety in the derivation is the fact that the broken symmetries are space-time symmetries, which leads to inverse Higgs constraints for the
components of the Maurer--Cartan form. As a check on our results, we also derived the effective action using the method of curvature invariants. 

Using this general effective action, we checked that the two-point function for the Goldstone $\pi$ and a fiducial weight-0 field $\chi$ correctly reproduces
known results~\cite{Rubakov:2009np,Creminelli:2010ba,Hinterbichler:2011qk}. We also computed the three-point correlation functions $\langle \chi\chi\chi\rangle$ and
$\langle \pi\chi\chi\rangle$, and checked their invariance under the unbroken $\mathfrak{so}(4,1)$ symmetries. As mentioned in the Introduction, the
correlation functions are also constrained by the full $\mathfrak{so}(4,2)$ symmetries, which should imply Ward identities relating $N$ and $N-1$ point correlators.
These novel constraints should in principle yield distinguishing relations that the smaller symmetry algebra of inflation cannot reproduce. 
The general effective action derived here offers a useful laboratory with which to identify and test such relations.

\vspace{-4pt}
It should be noted that while we have focused in this paper on the coset construction -- as it is best suited for the problem of constructing non-linear realizations -- there exist other powerful techniques for the construction of conformally-invariant actions. Perhaps the most elegant of these is the formalism of tractor caclulus. Most simply, tractors play the same role in conformal geometry that tensors play in Riemannian geometry. Tractor calculus was first introduced in \cite{gover1}, building on earlier ideas from the 1920's \cite{thomas1, thomas2}. Tractors live in $\mathbb R^{4,2}$, where the conformal group ${\rm SO}(4,2)$ acts naturally. A nice introduction to these ideas is given in \cite{eastwood}. Tractors provide a powerful formalism for handling conformal invariance; by contracting tractors and tractor covariant derivatives to construct scalars, one automatically obtains Weyl-invariant theories in four dimensions, analogous to how one ordinarily builds diffeomorphism-invariant actions with tensors. Tractor calculus has been applied to physical systems in many ways, most notably to address the origins of mass \cite{Gover:2008pt, Gover:2008sw} and to view Einstein gravity from a six-dimensional viewpoint \cite{Bonezzi:2010jr}. Although not included in our discussion, we have verified explicitly that the conformal actions constructed with apparatus of tractor calculus agree with those descending from the coset construction.

\vspace{-4pt}
Another method of constructing field theories with non-linearly realized symmetries is the embedded-brane technique of \cite{deRham:2010eu,Hinterbichler:2010xn,Goon:2011qf}, in which the physical space is imagined as a 3-brane floating in a non-dynamical bulk. The fields in the physical space-time then inherit non-linear symmetries from the Killing vectors of the higher-dimensional bulk. In \cite{Goon:2011qf}, this approach was used to construct effective field theories realizing various patterns of symmetry breaking to maximal subalgebras.

\vspace{-4pt}
{\bf Acknowledgments:} It is our pleasure to thank James Bonifacio, Paolo Creminelli, Garrett Goon, Lam Hui, Randy Kamien, Godfrey Miller, Alberto Nicolis, Riccardo Penco, Valery Rubakov, Leonardo Senatore, Mark Trodden and Andrew Waldron for helpful discussions.
This work is supported in part by funds from the University of Pennsylvania, NASA ATP grant NNX11AI95G and the Alfred P. Sloan Foundation.

\appendix
\section{A Six-Dimensional Perspective on the Negative Quartic Model}
\label{6Dpers}
Here we present the negative quartic model from a slightly new perspective. As was first shown by Dirac \cite{Dirac:1936fq}, the conformal group of $\mathbb R^{3,1}$ -- ${\rm SO}(4,2)$ -- has a natural action in six-dimensional Minkowski space with two time-like directions.  The ${\rm SO}(4,2)$ acts as the $6d$ Lorentz transformations which leave the six-dimensional light cone invariant. Related ideas appeared in \cite{Salam:1970qk, Fubini:1976jm} which consider spontaneous breaking of conformal symmetry. This ambient space construction can also be viewed from the tractor calculus viewpoint \cite{gover2, gover3,Gover:2009vc} and is also related to the two-times (2T) program \cite{Bars:2000qm}. A similar six-dimensional construction has also been profitable in constructing scattering amplitudes in $4d$ \cite{Weinberg:2010fx}. Here we follow mostly \cite{Fubini:1976jm} and interpret the results in a new way. We identify $\mathbb R^{3,1}$ as the intersection of a null hyperplane with the light cone in $\mathbb R^{4,2}$ via the embedding \cite{Fubini:1976jm}
\be
y^\mu = x^\mu~~~~~~~~~~~~~
y^5 = \frac{1-x_\mu x^\mu}{2}~~~~~~~~~~~~~
y^6 = \frac{1+x_\mu x^\mu}{2}~.
\label{6dcoords}
\ee
It is then possible to write the quartic conformally invariant model of \cite{Hinterbichler:2011qk} in six-dimensions (with the metric $\eta_{AB}={\rm diag}(\eta_{\mu\nu},1,-1)$) where the orthogonal group acts naturally
\be
S = \int\rd^6y\left[-\frac{1}{2}\eta^{AB}\partial_A\phi\partial_B\phi + \frac{\lambda}{4}\phi^4\right]~.
\label{6dp4}
\ee
The isometry algebra of this action is $\mathfrak{so}(4,2)$, generated by $\delta_{J_{AB}} = y_A\partial_B-y_B\partial_A$.
The relation between this six dimensional parameterization of the algebra and the standard four dimensional parameterization is given by (\ref{6dconfalg}).

In addition to the equation of motion, the dynamics should be independent of which null hyperplane we choose, that is the field $\phi$ must be invariant under scaling up and down the light cone, corresponding to dilations from the four-dimensional perspective. Requiring that the field transform with weight $1$ implies that it satisfies the equation \cite{Salam:1970qk, Fubini:1976jm}
\be
y^A\partial_A\phi + \phi = 0~.
\ee
The equation of motion for (\ref{6dp4}) is
\be
\square_6\phi -\lambda\phi^3 = 0~,
\ee
where $\square_6 = \eta^{AB}\partial_A\partial_B$. Both of these equations are solved by the field profile \cite{Fubini:1976jm}
\be
\bar\phi(y) = \sqrt{\frac{2}{\lambda}}\frac{1}{h_Ay^A}~,
\label{6dsoln}
\ee
where $h^A$ is a six-dimensional time-like unit vector. This background profile spontaneously breaks the conformal algebra down to the stabilizer of the vector $h^A$, in this case it is the de Sitter algebra $\mathfrak{so}(4,1)$,\footnote{Breaking to the anti de Sitter algebra or the Poincar\'e algebra can be achieved by taking space-like or null $h^A$, respectively \cite{Fubini:1976jm}.} we therefore have the symmetry breaking pattern
\be
\mathfrak{so}(4,2)\longrightarrow \mathfrak{so}(4,1)~.
\ee
It is also worth noting that the expression (\ref{6dsoln}) can be projected down to a solution of the four-dimensional equations of motion using the embedding (\ref{6dcoords}) -- the six-dimensional approach is merely a convenient way to make conformal symmetry manifest, much as Poincar\'e symmetry is manifest in the four dimensional theory. Having found these solutions to the equations of motion, we return to the four-dimensional picture and note that spectator fields will couple to the effective metric
\be
\bar g_{\mu\nu}^{\rm eff} = \bar\phi^2\eta_{\mu\nu} = \frac{2}{\lambda (h_Ay^A)^2}\eta_{\mu\nu}~.
\ee

We have not yet chosen an explicit $h^A$, which will correspond to an explicit coordinitization of de Sitter; in order to recover the parameterization considered above, consider the case where $h^A$ points along the $y^0 = x^0 \equiv t$ direction. This corresponds to the field profile (\ref{1otsoln}); it is relatively easy to check that the stabilizer of the vector $h^A$ is generated by
\be
\left\{\delta_{J_{5i}},~\delta_{J_{6i}},~\delta_{J_{ij}},~\delta_{D}\right\}~,
\ee
which form an $\mathfrak{so}(4,1)$ algebra and correspond precisely to the generators (\ref{1otgens}). In this case, the effective metric to which spectator fields couple is
\be
\bar g_{\mu\nu}^{\rm eff} = \frac{2}{\lambda t^2}\eta_{\mu\nu}~.
\ee
Alternatively, we could take $h^A$ to point along the  $y^6 = \frac{1}{2}(1+x_\mu x^\mu)$ direction, in which case the preserved de Sitter subalgebra is generated by
\be
\left\{\delta_{J_{5\mu}},~\delta_{J_{\mu\nu}}\right\}~.
\ee
and the effective de Sitter metric is
\be
\bar g^{\rm eff}_{\mu\nu} = \frac{8}{\lambda(1+x_\mu x^\mu)^2}\eta_{\mu\nu}~.
\ee


\begin{thebibliography}{99}

\bibitem{Starobinsky:1979ty} 
  A.~A.~Starobinsky,
  ``Relict Gravitation Radiation Spectrum and Initial State of the Universe. (In Russian),''
  JETP Lett.\  {\bf 30}, 682 (1979)
  [Pisma Zh.\ Eksp.\ Teor.\ Fiz.\  {\bf 30}, 719 (1979)].

\bibitem{Guth:1980zm} 
  A.~H.~Guth,
  ``The Inflationary Universe: A Possible Solution to the Horizon and Flatness Problems,''
  Phys.\ Rev.\ D {\bf 23}, 347 (1981).

\bibitem{Albrecht:1982wi} 
  A.~Albrecht and P.~J.~Steinhardt,
  ``Cosmology for Grand Unified Theories with Radiatively Induced Symmetry Breaking,''
  Phys.\ Rev.\ Lett.\  {\bf 48}, 1220 (1982).

\bibitem{Linde:1981mu} 
  A.~D.~Linde,
  ``A New Inflationary Universe Scenario: A Possible Solution of the Horizon, Flatness, Homogeneity, Isotropy and Primordial Monopole Problems,''
  Phys.\ Lett.\ B {\bf 108}, 389 (1982).

\bibitem{Gasperini:1992em}
  M.~Gasperini and G.~Veneziano,
  ``Pre - big bang in string cosmology,''
  Astropart.\ Phys.\  {\bf 1}, 317 (1993)
  \href{http://arxiv.org/abs/hep-th/9211021}{[arXiv:hep-th/9211021]}.

\bibitem{Gasperini:2002bn}
  M.~Gasperini and G.~Veneziano,
  ``The Pre - big bang scenario in string cosmology,''
  Phys.\ Rept.\  {\bf 373}, 1 (2003)
  \href{http://arxiv.org/abs/hep-th/0207130}{[arXiv:hep-th/0207130]}.

\bibitem{Gasperini:2007vw}
  M.~Gasperini and G.~Veneziano,
  ``String Theory and Pre-big bang Cosmology,''
  \href{http://arxiv.org/abs/hep-th/0703055}{arXiv:hep-th/0703055}.

\bibitem{Brandenberger:1988aj}
  R.~H.~Brandenberger and C.~Vafa,
  ``Superstrings in the Early Universe,''
  Nucl.\ Phys.\  B {\bf 316}, 391 (1989).

\bibitem{Nayeri:2005ck}
  A.~Nayeri, R.~H.~Brandenberger and C.~Vafa,
  ``Producing a scale-invariant spectrum of perturbations in a Hagedorn  phase
  of string cosmology,''
  Phys.\ Rev.\ Lett.\  {\bf 97}, 021302 (2006)
  \href{http://arxiv.org/abs/hep-th/0511140}{[arXiv:hep-th/0511140]}.

\bibitem{Brandenberger:2006xi}
  R.~H.~Brandenberger, A.~Nayeri, S.~P.~Patil and C.~Vafa,
  ``Tensor modes from a primordial Hagedorn phase of string cosmology,''
  Phys.\ Rev.\ Lett.\  {\bf 98}, 231302 (2007)
  \href{http://arxiv.org/abs/hep-th/0604126}{[arXiv:hep-th/0604126]}.
  
\bibitem{Brandenberger:2006vv}
  R.~H.~Brandenberger, A.~Nayeri, S.~P.~Patil and C.~Vafa,
  ``String gas cosmology and structure formation,''
  Int.\ J.\ Mod.\ Phys.\  A {\bf 22}, 3621 (2007)
  \href{http://arxiv.org/abs/hep-th/0608121}{[arXiv:hep-th/0608121]}.

\bibitem{Brandenberger:2006pr}
  R.~H.~Brandenberger {\it et al.},
  ``More on the Spectrum of Perturbations in String Gas Cosmology,''
  JCAP {\bf 0611}, 009 (2006)
  \href{http://arxiv.org/abs/hep-th/0608186}{[arXiv:hep-th/0608186]}.

\bibitem{Battefeld:2005av}
  T.~Battefeld and S.~Watson,
  ``String gas cosmology,''
  Rev.\ Mod.\ Phys.\  {\bf 78}, 435 (2006)
  \href{http://arxiv.org/abs/hep-th/0510022}{[arXiv:hep-th/0510022]}.

\bibitem{Khoury:2001wf}
  J.~Khoury, B.~A.~Ovrut, P.~J.~Steinhardt and N.~Turok,
  ``The ekpyrotic universe: Colliding branes and the origin of the hot big
  bang,''
  Phys.\ Rev.\  D {\bf 64}, 123522 (2001)
  \href{http://arxiv.org/abs/hep-th/0103239}{[arXiv:hep-th/0103239]}.

\bibitem{Donagi:2001fs}
  R.~Y.~Donagi, J.~Khoury, B.~A.~Ovrut, P.~J.~Steinhardt and N.~Turok,
  ``Visible branes with negative tension in heterotic M-theory,''
  JHEP {\bf 0111} (2001) 041
  \href{http://arxiv.org/abs/hep-th/0105199}{[arXiv:hep-th/0105199]}.

\bibitem{Khoury:2001bz}
  J.~Khoury, B.~A.~Ovrut, N.~Seiberg, P.~J.~Steinhardt and N.~Turok,
  ``From big crunch to big bang,''
  Phys.\ Rev.\  D {\bf 65}, 086007 (2002)
  \href{http://arxiv.org/abs/hep-th/0108187}{[arXiv:hep-th/0108187]}.

\bibitem{Khoury:2001zk}
  J.~Khoury, B.~A.~Ovrut, P.~J.~Steinhardt and N.~Turok,
  ``Density perturbations in the ekpyrotic scenario,''
  Phys.\ Rev.\  D {\bf 66}, 046005 (2002)
  \href{http://arxiv.org/abs/hep-th/0109050}{[arXiv:hep-th/0109050]}.

\bibitem{Lyth:2001pf}
  D.~H.~Lyth,
  ``The primordial curvature perturbation in the ekpyrotic universe,''
  Phys.\ Lett.\  B {\bf 524}, 1 (2002)
  \href{http://arxiv.org/abs/hep-ph/0106153}{[arXiv:hep-ph/0106153]}.

\bibitem{Brandenberger:2001bs}
  R.~Brandenberger and F.~Finelli,
  ``On the spectrum of fluctuations in an effective field theory of the
  ekpyrotic universe,''
  JHEP {\bf 0111}, 056 (2001)
  \href{http://arxiv.org/abs/hep-th/0109004}{[arXiv:hep-th/0109004]}.

\bibitem{Steinhardt:2001st}
  P.~J.~Steinhardt and N.~Turok,
  ``Cosmic evolution in a cyclic universe,''
  Phys.\ Rev.\  D {\bf 65}, 126003 (2002)
  \href{http://arxiv.org/abs/hep-th/0111098}{[arXiv:hep-th/0111098]}.

\bibitem{Notari:2002yc}
  A.~Notari and A.~Riotto,
  ``Isocurvature perturbations in the ekpyrotic universe,''
  Nucl.\ Phys.\  B {\bf 644}, 371 (2002)
  \href{http://arxiv.org/abs/hep-th/0205019}{[arXiv:hep-th/0205019]}.

\bibitem{Finelli:2002we}
  F.~Finelli,
  ``Assisted contraction,''
  Phys.\ Lett.\  B {\bf 545}, 1 (2002)
  \href{http://arxiv.org/abs/hep-th/0206112}{[arXiv:hep-th/0206112]}.

\bibitem{Tsujikawa:2002qc}
  S.~Tsujikawa, R.~Brandenberger and F.~Finelli,
  ``On the construction of nonsingular pre-big-bang and ekpyrotic cosmologies
  and the resulting density perturbations,''
  Phys.\ Rev.\  D {\bf 66}, 083513 (2002)
  \href{http://arxiv.org/abs/hep-th/0207228}{[arXiv:hep-th/0207228]}.
 
\bibitem{Gratton:2003pe}
  S.~Gratton, J.~Khoury, P.~J.~Steinhardt and N.~Turok,
  ``Conditions for generating scale-invariant density perturbations,''
  Phys.\ Rev.\  D {\bf 69}, 103505 (2004)
  \href{http://arxiv.org/abs/astro-ph/0301395}{[arXiv:astro-ph/0301395]}.

\bibitem{Tolley:2003nx}
  A.~J.~Tolley, N.~Turok and P.~J.~Steinhardt,
  ``Cosmological perturbations in a big crunch / big bang space-time,''
  Phys.\ Rev.\  D {\bf 69}, 106005 (2004)
  \href{http://arxiv.org/abs/hep-th/0306109}{[arXiv:hep-th/0306109]}.
  
\bibitem{Craps:2003ai}
  B.~Craps and B.~A.~Ovrut,
  ``Global fluctuation spectra in big crunch / big bang string vacua,''
  Phys.\ Rev.\  D {\bf 69}, 066001 (2004)
  \href{http://arxiv.org/abs/hep-th/0308057}{[arXiv:hep-th/0308057]}.

\bibitem{Khoury:2003vb}
  J.~Khoury, P.~J.~Steinhardt and N.~Turok,
  ``Great expectations: Inflation versus cyclic predictions for spectral
  tilt,''
  Phys.\ Rev.\ Lett.\  {\bf 91}, 161301 (2003)
   \href{http://arxiv.org/abs/astro-ph/0302012}{[arXiv:astro-ph/0302012]}.

\bibitem{Khoury:2003rt}
  J.~Khoury, P.~J.~Steinhardt and N.~Turok,
  ``Designing Cyclic Universe Models,''
  Phys.\ Rev.\ Lett.\  {\bf 92}, 031302 (2004)
   \href{http://arxiv.org/abs/hep-th/0307132}{[arXiv:hep-th/0307132]}.

\bibitem{Khoury:2004xi}
  J.~Khoury,
  ``A briefing on the ekpyrotic / cyclic universe,''
   \href{http://arxiv.org/abs/astro-ph/0401579}{arXiv:astro-ph/0401579}.

\bibitem{Creminelli:2004jg}
  P.~Creminelli, A.~Nicolis and M.~Zaldarriaga,
  ``Perturbations in bouncing cosmologies: Dynamical attractor vs scale
  invariance,''
  Phys.\ Rev.\  D {\bf 71}, 063505 (2005)
   \href{http://arxiv.org/abs/hep-th/0411270}{[arXiv:hep-th/0411270]}.

\bibitem{Lehners:2007ac}
  J.~L.~Lehners, P.~McFadden, N.~Turok and P.~J.~Steinhardt,
  ``Generating ekpyrotic curvature perturbations before the big bang,''
  Phys.\ Rev.\  D {\bf 76}, 103501 (2007)
   \href{http://arxiv.org/abs/hep-th/0702153}{[arXiv:hep-th/0702153]}.

\bibitem{Buchbinder:2007ad}
  E.~I.~Buchbinder, J.~Khoury and B.~A.~Ovrut,
  ``New Ekpyrotic Cosmology,''
  Phys.\ Rev.\  D {\bf 76}, 123503 (2007)
   \href{http://arxiv.org/abs/hep-th/0702154}{[arXiv:hep-th/0702154]}.

\bibitem{Creminelli:2007aq}
  P.~Creminelli and L.~Senatore,
  ``A smooth bouncing cosmology with scale invariant spectrum,''
  JCAP {\bf 0711}, 010 (2007)
   \href{http://arxiv.org/abs/hep-th/0702165}{[arXiv:hep-th/0702165]}.

\bibitem{Buchbinder:2007tw}
  E.~I.~Buchbinder, J.~Khoury and B.~A.~Ovrut,
  ``On the Initial Conditions in New Ekpyrotic Cosmology,''
  JHEP {\bf 0711}, 076 (2007)
   \href{http://arxiv.org/abs/0706.3903}{[arXiv:0706.3903 [hep-th]]}.

\bibitem{Buchbinder:2007at}
  E.~I.~Buchbinder, J.~Khoury and B.~A.~Ovrut,
  ``Non-Gaussianities in New Ekpyrotic Cosmology,''
  Phys.\ Rev.\ Lett.\  {\bf 100}, 171302 (2008)
  \href{http://arxiv.org/abs/0710.5172}{[arXiv:0710.5172 [hep-th]]}.

\bibitem{Koyama:2007mg}
  K.~Koyama and D.~Wands,
  ``Ekpyrotic collapse with multiple fields,''
  JCAP {\bf 0704}, 008 (2007)
  \href{http://arxiv.org/abs/hep-th/0703040}{[arXiv:hep-th/0703040]}.

\bibitem{Koyama:2007ag}
  K.~Koyama, S.~Mizuno and D.~Wands,
  ``Curvature perturbations from ekpyrotic collapse with multiple fields,''
  Class.\ Quant.\ Grav.\  {\bf 24}, 3919 (2007)
  \href{http://arxiv.org/abs/0704.1152}{[arXiv:0704.1152 [hep-th]]}.

\bibitem{Lehners:2007wc}
  J.~L.~Lehners and P.~J.~Steinhardt,
  ``Non-Gaussian Density Fluctuations from Entropically Generated Curvature
  Perturbations in Ekpyrotic Models,''
  Phys.\ Rev.\  D {\bf 77}, 063533 (2008)
  [Erratum-ibid.\  D {\bf 79}, 129903 (2009)]
  \href{http://arxiv.org/abs/0712.3779}{[arXiv:0712.3779 [hep-th]]}.

\bibitem{Lehners:2008my}
  J.~L.~Lehners and P.~J.~Steinhardt,
  ``Intuitive understanding of non-gaussianity in ekpyrotic and cyclic
  models,''
  Phys.\ Rev.\  D {\bf 78}, 023506 (2008)
  [Erratum-ibid.\  D {\bf 79}, 129902 (2009)]
  \href{http://arxiv.org/abs/0804.1293}{[arXiv:0804.1293 [hep-th]]}.

\bibitem{Lehners:2009qu}
  J.~L.~Lehners and P.~J.~Steinhardt,
  ``Non-Gaussianity Generated by the Entropic Mechanism in Bouncing Cosmologies
  Made Simple,''
  Phys.\ Rev.\  D {\bf 80}, 103520 (2009)
  \href{http://arxiv.org/abs/0909.2558}{[arXiv:0909.2558 [hep-th]]}.

\bibitem{Khoury:2009my} 
  J.~Khoury and P.~J.~Steinhardt,
  ``Adiabatic Ekpyrosis: Scale-Invariant Curvature Perturbations from a Single Scalar Field in a Contracting Universe,''
  Phys.\ Rev.\ Lett.\  {\bf 104}, 091301 (2010)
  \href{http://arxiv.org/abs/0910.2230}{[arXiv:0910.2230 [hep-th]]}.
 
\bibitem{Khoury:2011ii} 
  J.~Khoury and P.~J.~Steinhardt,
  ``Generating Scale-Invariant Perturbations from Rapidly-Evolving Equation of State,''
  Phys.\ Rev.\ D {\bf 83}, 123502 (2011)
  \href{http://arxiv.org/abs/1101.3548}{[arXiv:1101.3548 [hep-th]]}.

\bibitem{Joyce:2011ta} 
  A.~Joyce and J.~Khoury,
  ``Scale Invariance via a Phase of Slow Expansion,''
  Phys.\ Rev.\ D {\bf 84}, 023508 (2011)
  \href{http://arxiv.org/abs/1104.4347}{[arXiv:1104.4347 [hep-th]]}.

\bibitem{Khoury:2010gw} 
  J.~Khoury and G.~E.~J.~Miller,
  ``Towards a Cosmological Dual to Inflation,''
  Phys.\ Rev.\ D {\bf 84}, 023511 (2011)
  \href{http://arxiv.org/abs/1012.0846}{[arXiv:1012.0846 [hep-th]]}.

\bibitem{Joyce:2011kh} 
  A.~Joyce and J.~Khoury,
  ``Strong Coupling Problem with Time-Varying Sound Speed,''
  Phys.\ Rev.\ D {\bf 84}, 083514 (2011)
  \href{http://arxiv.org/abs/1107.3550}{[arXiv:1107.3550 [hep-th]]}.

\bibitem{Baumann:2011dt} 
  D.~Baumann, L.~Senatore and M.~Zaldarriaga,
  ``Scale-Invariance and the Strong Coupling Problem,''
  JCAP {\bf 1105}, 004 (2011)
  \href{http://arxiv.org/abs/1101.3320}{[arXiv:1101.3320 [hep-th]]}.

\bibitem{Geshnizjani:2011dk} 
  G.~Geshnizjani, W.~H.~Kinney and A.~M.~Dizgah,
  ``General conditions for scale-invariant perturbations in an expanding universe,''
  JCAP {\bf 1111}, 049 (2011)
  \href{http://arxiv.org/abs/1107.1241}{[arXiv:1107.1241 [astro-ph.CO]]}.

\bibitem{Wands:1998yp} 
  D.~Wands,
  ``Duality invariance of cosmological perturbation spectra,''
  Phys.\ Rev.\ D {\bf 60}, 023507 (1999)
  \href{http://arxiv.org/abs/gr-qc/9809062}{[gr-qc/9809062]}.

\bibitem{Finelli:2001sr} 
  F.~Finelli and R.~Brandenberger,
  ``On the generation of a scale invariant spectrum of adiabatic fluctuations in cosmological models with a contracting phase,''
  Phys.\ Rev.\ D {\bf 65}, 103522 (2002)
  \href{http://arxiv.org/abs/hep-th/0112249}{[hep-th/0112249]}.

\bibitem{Rubakov:2009np} 
  V.~A.~Rubakov,
  ``Harrison-Zeldovich spectrum from conformal invariance,''
  JCAP {\bf 0909}, 030 (2009)
  \href{http://arxiv.org/abs/0906.3693}{[arXiv:0906.3693 [hep-th]]}.

\bibitem{Creminelli:2010ba} 
  P.~Creminelli, A.~Nicolis and E.~Trincherini,
  ``Galilean Genesis: An Alternative to inflation,''
  JCAP {\bf 1011}, 021 (2010)
  \href{http://arxiv.org/abs/1007.0027}{[arXiv:1007.0027 [hep-th]]}.

\bibitem{Hinterbichler:2011qk} 
  K.~Hinterbichler and J.~Khoury,
  ``The Pseudo-Conformal Universe: Scale Invariance from Spontaneous Breaking of Conformal Symmetry,''
  JCAP {\bf 1204}, 023 (2012)
  \href{http://arxiv.org/abs/1106.1428}{[arXiv:1106.1428 [hep-th]]}.

\bibitem{Lyth:2001nq} 
  D.~H.~Lyth and D.~Wands,
  ``Generating the curvature perturbation without an inflaton,''
  Phys.\ Lett.\ B {\bf 524}, 5 (2002)
  \href{http://arxiv.org/abs/hep-ph/0110002}{[hep-ph/0110002]}.

\bibitem{Kofman:2003nx} 
  L.~Kofman,
  ``Probing string theory with modulated cosmological fluctuations,''
  \href{http://arxiv.org/abs/astro-ph/0303614}{astro-ph/0303614}.

\bibitem{Dvali:2003em} 
  G.~Dvali, A.~Gruzinov and M.~Zaldarriaga,
  ``A new mechanism for generating density perturbations from inflation,''
  Phys.\ Rev.\ D {\bf 69}, 023505 (2004)
  \href{http://arxiv.org/abs/astro-ph/0303591}{[astro-ph/0303591]}.

\bibitem{Craps:2007ch} 
  B.~Craps, T.~Hertog and N.~Turok,
  ``Quantum Resolution of Cosmological Singularities using AdS/CFT,''
  \href{http://arxiv.org/abs/0712.4180}{arXiv:0712.4180 [hep-th]}.

\bibitem{Coleman:1969sm} 
  S.~R.~Coleman, J.~Wess and B.~Zumino,
  ``Structure of phenomenological Lagrangians. 1.,''
  Phys.\ Rev.\  {\bf 177}, 2239 (1969).

\bibitem{Callan:1969sn} 
  C.~G.~Callan, Jr., S.~R.~Coleman, J.~Wess and B.~Zumino,
  ``Structure of phenomenological Lagrangians. 2.,''
  Phys.\ Rev.\  {\bf 177}, 2247 (1969).

\bibitem{volkov}
D.~V.~Volkov, 
``Phenomenological Lagrangians,"
Sov.\ J.\ Particles\ Nucl. {\bf 4}, 3 (1973).

\bibitem{Low:2001bw} 
  I.~Low and A.~V.~Manohar,
  ``Spontaneously broken space-time symmetries and Goldstone's theorem,''
  Phys.\ Rev.\ Lett.\  {\bf 88}, 101602 (2002)
  \href{http://arxiv.org/abs/hep-th/0110285}{[hep-th/0110285]}.

\bibitem{Ivanov:1975zq} 
  E.~A.~Ivanov and V.~I.~Ogievetsky,
  ``The Inverse Higgs Phenomenon in Nonlinear Realizations,''
  Teor.\ Mat.\ Fiz.\  {\bf 25}, 164 (1975).

\bibitem{Cheung:2007st} 
  C.~Cheung, P.~Creminelli, A.~L.~Fitzpatrick, J.~Kaplan and L.~Senatore,
  ``The Effective Field Theory of Inflation,''
  JHEP {\bf 0803}, 014 (2008)
  \href{http://arxiv.org/abs/0709.0293}{[arXiv:0709.0293 [hep-th]]}.

\bibitem{DiFrancesco:1997nk} 
  P.~Di Francesco, P.~Mathieu and D.~Senechal,
  ``Conformal field theory,''
  New York, USA: Springer (1997) 890 p

\bibitem{Antoniadis:2011ib} 
  I.~Antoniadis, P.~O.~Mazur and E.~Mottola,
  ``Conformal Invariance, Dark Energy, and CMB Non-Gaussianity,''
 \href{http://arxiv.org/abs/1103.4164}{arXiv:1103.4164 [gr-qc]}.

\bibitem{Creminelli:2011mw} 
  P.~Creminelli,
  ``Conformal invariance of scalar perturbations in inflation,''
  Phys.\ Rev.\ D {\bf 85}, 041302 (2012)
  \href{http://arxiv.org/abs/1108.0874}{[arXiv:1108.0874 [hep-th]]}.

\bibitem{Maldacena:2011nz} 
  J.~M.~Maldacena and G.~L.~Pimentel,
  ``On graviton non-Gaussianities during inflation,''
  JHEP {\bf 1109}, 045 (2011)
  \href{http://arxiv.org/abs/1104.2846}{[arXiv:1104.2846 [hep-th]]}.

\bibitem{Nicolis:2008in} 
  A.~Nicolis, R.~Rattazzi and E.~Trincherini,
  ``The Galileon as a local modification of gravity,''
  Phys.\ Rev.\ D {\bf 79}, 064036 (2009)
  \href{http://arxiv.org/abs/0811.2197}{[arXiv:0811.2197 [hep-th]]}.

\bibitem{Osipov:2010ee} 
  M.~Osipov and V.~Rubakov,
  ``Scalar tilt from broken conformal invariance,''
  JETP Lett.\  {\bf 93}, 52 (2011)
  \href{http://arxiv.org/abs/1007.3417}{[arXiv:1007.3417 [hep-th]]}.

\bibitem{Libanov:2010nk} 
  M.~Libanov and V.~Rubakov,
  ``Cosmological density perturbations from conformal scalar field: infrared properties and statistical anisotropy,''
  JCAP {\bf 1011}, 045 (2010)
  \href{http://arxiv.org/abs/1007.4949}{[arXiv:1007.4949 [hep-th]]}.

\bibitem{Libanov:2010ci} 
  M.~Libanov, S.~Mironov and V.~Rubakov,
  ``Properties of scalar perturbations generated by conformal scalar field,''
  Prog.\ Theor.\ Phys.\ Suppl.\  {\bf 190}, 120 (2011)
  \href{http://arxiv.org/abs/1012.5737}{[arXiv:1012.5737 [hep-th]]}.

\bibitem{Libanov:2011bk} 
  M.~Libanov, S.~Mironov and V.~Rubakov,
  ``Non-Gaussianity of scalar perturbations generated by conformal mechanisms,''
  Phys.\ Rev.\ D {\bf 84}, 083502 (2011)
  \href{http://arxiv.org/abs/1105.6230}{[arXiv:1105.6230 [astro-ph.CO]]}.

\bibitem{Libanov:2011zy} 
  M.~Libanov and V.~Rubakov,
  ``Dynamical vs spectator models of (pseudo-)conformal Universe,''
  \href{http://arxiv.org/abs/1107.1036}{arXiv:1107.1036 [hep-th]}.

\bibitem{Piao:2011bz} 
  Y.~-S.~Piao,
  ``Conformally Dual to Inflation,''
  \href{http://arxiv.org/abs/1112.3737}{arXiv:1112.3737 [hep-th]}.

\bibitem{Piao:2011mq} 
  Y.~-S.~Piao,
  ``Gravitational Wave During Slowly Evolving,''
  \href{http://arxiv.org/abs/1109.4266}{arXiv:1109.4266 [hep-th]}.

\bibitem{Weinberg:1968de} 
  S.~Weinberg,
  ``Nonlinear realizations of chiral symmetry,''
  Phys.\ Rev.\  {\bf 166}, 1568 (1968).

\bibitem{xthschool}
 V.~I.~Ogievetsky,
 ``Nonlinear Realizations of Internal and Space-time Symmetries,"
 Proc. of X--th Winter School of Theoretical Physics in Karpacz, Vol. 1, Wroclaw 227 (1974)

\bibitem{Zumino:1970tu} 
  B.~Zumino,
  ``Effective Lagrangians and broken symmetries,''
  In *Brandeis Univ. 1970, Lectures On Elementary Particles And Quantum Field Theory, Vol. 2*, Cambridge, Mass. 1970, 437-500.

\bibitem{Bellucci:2002ji} 
  S.~Bellucci, E.~Ivanov and S.~Krivonos,
  ``AdS / CFT equivalence transformation,''
  Phys.\ Rev.\ D {\bf 66}, 086001 (2002)
  [Erratum-ibid.\ D {\bf 67}, 049901 (2003)]
  \href{http://arxiv.org/abs/hep-th/0206126}{[hep-th/0206126]}.

\bibitem{Salam:1970qk} 
  A.~Salam and J.~A.~Strathdee,
  ``Nonlinear realizations. 2. Conformal symmetry,''
  Phys.\ Rev.\  {\bf 184}, 1760 (1969).

\bibitem{Isham:1970xz} 
  C.~J.~Isham, A.~Salam and J.~A.~Strathdee,
  ``Broken chiral and conformal symmetry in an effective-lagrangian formalism,''
  Phys.\ Rev.\ D {\bf 2}, 685 (1970).

\bibitem{Isham:1970gz} 
  C.~J.~Isham, A.~Salam and J.~A.~Strathdee,
  ``Spontaneous breakdown of conformal symmetry,''
  Phys.\ Lett.\ B {\bf 31}, 300 (1970).

\bibitem{McArthur:2010zm} 
  I.~N.~McArthur,
  ``Nonlinear realizations of symmetries and unphysical Goldstone bosons,''
  JHEP {\bf 1011}, 140 (2010)
  \href{http://arxiv.org/abs/1009.3696}{[arXiv:1009.3696 [hep-th]]}.

\bibitem{Clark:2005ht} 
  T.~E.~Clark, S.~T.~Love, M.~Nitta and T.~ter Veldhuis,
  ``${\rm AdS}_{d+1} \to {\rm AdS}_d$,''
  J.\ Math.\ Phys.\  {\bf 46}, 102304 (2005)
  \href{http://arxiv.org/abs/hep-th/0501241}{[hep-th/0501241]}.

\bibitem{Goon:2012dy} 
  G.~Goon, K.~Hinterbichler, A.~Joyce and M.~Trodden,
  ``Galileons as Wess-Zumino Terms,''
  \href{http://arxiv.org/abs/1203.3191}{arXiv:1203.3191 [hep-th]}.

\bibitem{Maldacena:2002vr} 
  J.~M.~Maldacena,
  ``Non-Gaussian features of primordial fluctuations in single field inflationary models,''
  JHEP {\bf 0305}, 013 (2003)
  \href{http://arxiv.org/abs/astro-ph/0210603}{[astro-ph/0210603]}.

\bibitem{Weinberg:2005vy} 
  S.~Weinberg,
  ``Quantum contributions to cosmological correlations,''
  Phys.\ Rev.\ D {\bf 72}, 043514 (2005)
  \href{http://arxiv.org/abs/hep-th/0506236}{[hep-th/0506236]}.

\bibitem{Zaldarriaga:2003my} 
  M.~Zaldarriaga,
  ``Non-Gaussianities in models with a varying inflaton decay rate,''
  Phys.\ Rev.\ D {\bf 69}, 043508 (2004)
  \href{http://arxiv.org/abs/astro-ph/0306006}{[astro-ph/0306006]}.

\bibitem{gover1}
T.~N.~Bailey, M.~G.~Eastwood and A.~R.~Gover,\
 ``Thomas's Structure Bundle for Conformal, Projective and Related Structures,"\
Rocky Mountain Journal of Mathematics {\bf 24} (1994)

\bibitem{thomas1}
T.~Y.~Thomas,\
``On Conformal Geometry,"\
Proc. Nat. Acad. Sci. {\bf 12}, 352-359 (1926)

\bibitem{thomas2}
T.~Y.~Thomas,\
 ``Conformal Tensors,"\
Proc. Nat. Acad. Sci. {\bf 18}, 103-112 (1932)

\bibitem{eastwood}
M.~Eastwood,\
 ``Notes on Conformal Differential Geometry,"\
 Suppl. Rendi. Circ. Mat. Palermo {\bf 43}, 57-76 (1996)

\bibitem{Gover:2008pt} 
  A.~R.~Gover, A.~Shaukat and A.~Waldron,
  ``Weyl Invariance and the Origins of Mass,''
  Phys.\ Lett.\ B {\bf 675}, 93 (2009)
  \href{http://arxiv.org/abs/0812.3364}{[arXiv:0812.3364 [hep-th]]}.

\bibitem{Gover:2008sw} 
  A.~R.~Gover, A.~Shaukat and A.~Waldron,
  ``Tractors, Mass and Weyl Invariance,''
  Nucl.\ Phys.\ B {\bf 812}, 424 (2009)
  \href{http://arxiv.org/abs/0810.2867}{[arXiv:0810.2867 [hep-th]]}.

\bibitem{Bonezzi:2010jr} 
  R.~Bonezzi, E.~Latini and A.~Waldron,
  ``Gravity, Two Times, Tractors, Weyl Invariance and Six Dimensional Quantum Mechanics,''
  Phys.\ Rev.\ D {\bf 82}, 064037 (2010)
  \href{http://arxiv.org/abs/1007.1724}{[arXiv:1007.1724 [hep-th]]}.

\bibitem{deRham:2010eu} 
  C.~de Rham and A.~J.~Tolley,
  ``DBI and the Galileon reunited,''
  JCAP {\bf 1005}, 015 (2010)
  \href{http://arxiv.org/abs/1003.5917}{[arXiv:1003.5917 [hep-th]]}.

\bibitem{Hinterbichler:2010xn} 
  K.~Hinterbichler, M.~Trodden and D.~Wesley,
  ``Multi-field galileons and higher co-dimension branes,''
  Phys.\ Rev.\ D {\bf 82}, 124018 (2010)
  \href{http://arxiv.org/abs/1008.1305}{[arXiv:1008.1305 [hep-th]]}.

\bibitem{Goon:2011qf} 
  G.~Goon, K.~Hinterbichler and M.~Trodden,
  ``Symmetries for Galileons and DBI scalars on curved space,''
  JCAP {\bf 1107}, 017 (2011)
  \href{http://arxiv.org/abs/1103.5745}{[arXiv:1103.5745 [hep-th]]}.

\bibitem{Dirac:1936fq} 
  P.~A.~M.~Dirac,
  ``Wave equations in conformal space,''
  Annals Math.\  {\bf 37}, 429 (1936).

\bibitem{Fubini:1976jm} 
  S.~Fubini,
  ``A New Approach to Conformal Invariant Field Theories,''
  Nuovo Cim.\ A {\bf 34}, 521 (1976).

\bibitem{gover2}
A.~R.~Gover, L.~J.~Peterson, Lawrence, 
``The ambient obstruction tensor and the conformal deformation complex," 
Pacific J. Math. 226 (2006) no. 2, 309-351

\bibitem{gover3}
 A.~\v{C}ap, and A.~R.~Gover, 
``Standard tractors and the conformal ambient metric construction,"
Ann. Global Anal. Geom. 24 (2003), 231-295.

\bibitem{Gover:2009vc} 
  A.~R.~Gover and A.~Waldron,
  ``The so(d+2,2) Minimal Representation and Ambient Tractors: the Conformal Geometry of Momentum Space,''
  \href{http://arxiv.org/abs/0903.1394}{arXiv:0903.1394 [hep-th]}.

\bibitem{Bars:2000qm} 
  I.~Bars,
  ``Survey of two time physics,''
  Class.\ Quant.\ Grav.\  {\bf 18}, 3113 (2001)
  \href{http://arxiv.org/abs/hep-th/0008164}{[hep-th/0008164]}.

\bibitem{Weinberg:2010fx} 
  S.~Weinberg,
  ``Six-dimensional Methods for Four-dimensional Conformal Field Theories,''
  Phys.\ Rev.\ D {\bf 82}, 045031 (2010)
  \href{http://arxiv.org/abs/1006.3480}{[arXiv:1006.3480 [hep-th]]}.






\end{thebibliography}
\end{document}